\documentclass[aps,prx,twocolumn,superscriptaddress,showpacs,preprintnumbers,longbibliography]{revtex4-2}
\usepackage[colorlinks,bookmarks=false,citecolor=blue,linkcolor=red,urlcolor=blue]{hyperref}
\usepackage{physics}
\input pdfcolor.tex
\usepackage{colortbl,amsthm,txfonts}
\usepackage{verbatim}
\usepackage{graphicx}
\usepackage{epsfig}
\usepackage{dcolumn}
\usepackage{bm}

\usepackage{amsmath,amssymb}

\makeatletter
\renewcommand*\env@matrix[1][*\c@MaxMatrixCols c]{%                                                                   
  
  \hskip -\arraycolsep
  \let\@ifnextchar\new@ifnextchar
  \array{#1}}
\makeatother

\usepackage{appendix}

\begin{document}

\title{%2D Anderson Localization and KPZ Physics:\\ 
%Categorizing Conductance Fluctuations into the Tracy-Widom GOE and GUE Universality Classes
2D Anderson Localization and KPZ sub-Universality Classes:\\
sensitivity to boundary conditions and insensitivity to symmetry classes
}

\author{Nyayabanta Swain}
\email{nyayabanta@nus.edu.sg}
\affiliation{MajuLab, CNRS-UCA-SU-NUS-NTU International Joint Research Unit IRL 3654, Singapore}
\affiliation{Centre for Quantum Technologies, National University of Singapore, Singapore 117543}

\author{Shaffique Adam}
\affiliation{Department of Materials Science and Engineering, National University of Singapore, 9 Engineering Drive 1, Singapore 117575}
\affiliation{Department of Physics, Washington University in St. Louis, St. Louis, Missouri 63130, United States}

\author{\mbox{Gabriel Lemari\'e}}
\email{lemarie@irsamc.ups-tlse.fr}
\affiliation{MajuLab, CNRS-UCA-SU-NUS-NTU International Joint Research Unit IRL 3654, Singapore}
\affiliation{Centre for Quantum Technologies, National University of Singapore, Singapore 117543}
\affiliation{Laboratoire de Physique Th\'{e}orique, Universit\'{e} de Toulouse, CNRS, UPS, France}

\begin{abstract}
We challenge two foundational principles of localization physics by analyzing conductance fluctuations in two dimensions with unprecedented precision: (i) the Thouless criterion, which defines localization as insensitivity to boundary conditions, and (ii) that symmetry determines the universality class of Anderson localization. We reveal that the fluctuations of the conductance logarithm fall into distinct sub-universality classes inherited from Kardar-Parisi-Zhang (KPZ) physics, dictated by the lead configurations of the scattering system and unaffected by the presence of a magnetic field. Distinguishing between these probability distributions poses a significant challenge due to their striking similarity, requiring sampling beyond the usual threshold of $\sim 10^{-6}$ accessible through independent disorder realizations. To overcome this, we implement an importance-sampling scheme—a Monte Carlo approach in disorder space—that enables us to probe rare disorder configurations and sample probability distribution tails down to $10^{-30}$. This unprecedented precision allows us to unambiguously differentiate between KPZ sub-universality classes of conductance fluctuations for different lead configurations, while demonstrating the insensitivity to magnetic fields.  
\end{abstract}

\date{\today}
\maketitle

\section{Introduction}

Anderson localization—the absence of diffusive transport for a wave that is coherently scattered multiple times by a disordered medium—was discovered over 60 years ago \cite{Anderson:PR58}. Since then, extensive theoretical and experimental studies across diverse systems, from classical waves to matter waves, might suggest a thorough understanding of this phenomenon \cite{Evers_RMP_2008, abrahams2010, lagendijk2009fifty, sanchez2010disordered, santhanam2022quantum}. Attention has shifted to its extension in the presence of many-body interactions, prompting a fascinating re-evaluation of statistical physics \cite{abanin2018ergodicity, alet2017many, sierant2024many}. However, significant open questions remain in Anderson localization. Even just above the exactly solved one-dimensional case \cite{RevModPhys.69.731, Mirlin_PR_2000}, in two dimensions (2D) and beyond, our understanding relies on approximations valid only in regimes of weak disorder \cite{Evers_RMP_2008, abrahams2010}. For instance, it is still unclear whether our understanding of the Anderson transition holds in high dimensionality, where the phenomenon occurs at large disorder levels \cite{PhysRevB.95.094204, zirnbauer2023wegner, altshuler2024renormal}.

In this work, we challenge two foundational principles of localization physics by examining the still unresolved case of 2D Anderson localization with unprecedented precision (see \cite{PhysRevB.46.9984, prior2005conductance, Somoza_2007, prior2009conductance, Somoza_2015} for pioneering works). First, a key criterion distinguishing the diffusive regime from localization is the Thouless criterion, which posits that the localized phase is insensitive to changes in boundary conditions \cite{edwards1972numerical}. The corresponding Thouless energy—defined as the energy shift of a state when changing from periodic to anti-periodic boundary conditions—and the Thouless conductance, calculated as the ratio of Thouless energy to mean level spacing, are central to the scaling theory of localization \cite{abrahams1979}. Second, symmetry classes play a crucial role: the critical properties of the Anderson transition vary based on factors like time-reversal symmetry, magnetic fields, spin-orbit scattering, and topological characteristics \cite{Evers_RMP_2008}. This insight led to the celebrated 10-fold way of symmetry classes \cite{PhysRevB.55.1142, Evers_RMP_2008}.

Anderson’s assertion that “no real atom is an average atom” \cite{anderson1978local} highlights the importance of accounting for fluctuations to characterize disorder-induced localization. In this paper, we demonstrate that, in two dimensions, conductance fluctuations follow a universal distribution that depends crucially on boundary conditions while remaining unaffected by the symmetry class. This contradiction of the two principles outlined above arises from the Kardar-Parisi-Zhang (KPZ) physics that governs the strongly localized regime. KPZ physics~\cite{KPZ_1986,Corwin_2012,KPZ_30yrs_2015,Takeuchi_2018,Spohn_2020} provides a universal understanding of various classical processes, including the growth of rough interfaces and directed polymers in random media~\cite{HALPINHEALY1995,Johansson2000,Prahofer_2000_PRL, PhysRevLett.129.260603}. Surprisingly, recent studies have also identified KPZ properties in quantum systems, such as one-dimensional magnets~\cite{KPZ_Heisenberg_magnet_2019,KPZ_expt_spin_chain_2021,KPZ_expt_1D_chain_2022}, driven-dissipative quantum fluids~\cite{KPZ_expt_polariton_2022}, random unitary circuits~\cite{Nahum_2017}, and, central to this paper, Anderson localization in two dimensions~\cite{PhysRevB.46.9984, prior2005conductance, Somoza_2007, prior2009conductance, Somoza_2015, Lemarie_PRL_2019,musen_2023}.

In the realm of Anderson localization, universal fluctuations play a pivotal role. They manifest in various forms, such as universal conductance fluctuations~\cite{PhysRevB.35.1039,PhysRevLett.60.1089,prior2005conductance,prior2009conductance}, random matrix statistics~\cite{RMT_transport_Beenakker_1997,Evers_RMP_2008}, log-normal distributions in one-dimensional localized systems~\cite{RMT_transport_Beenakker_1997,Mirlin_PR_2000}, multifractal statistics in the critical regime of the Anderson transition~\cite{Lee_RMP_1985,castellani1986multifractal,Evers_RMP_2008,feigel2010fractal}, and KPZ physics in two dimension \cite{Somoza_2007,Lemarie_PRL_2019,musen_2023}, impacting a wide range of observables.

One of the most valuable aspects of drawing an analogy with KPZ physics is that the rich mathematical results in this field can reveal new laws and behaviors in the original system \cite{HALPINHEALY1995,Kriecherbauer_2010,Corwin_2012,KPZ_30yrs_2015,Quastel_Spohn_KPZ_rev_2015,Sasamoto_1D_KPZ_rev_2016}. In the context of Anderson localization, the analogy with KPZ physics predicts a set of properties that are not only counterintuitive and nearly impossible to foresee but also remarkably precise, as we will demonstrate.

%%%%%%%%%%%%%%%%%%%%%%%%%%%%%%%%%%%%%%%%%%%%%%%%%%%%%%%%%%%%%%%%%%%%%%%%%%%
\begin{figure}[t]
\begin{center}
\includegraphics[width=\linewidth, angle=0]{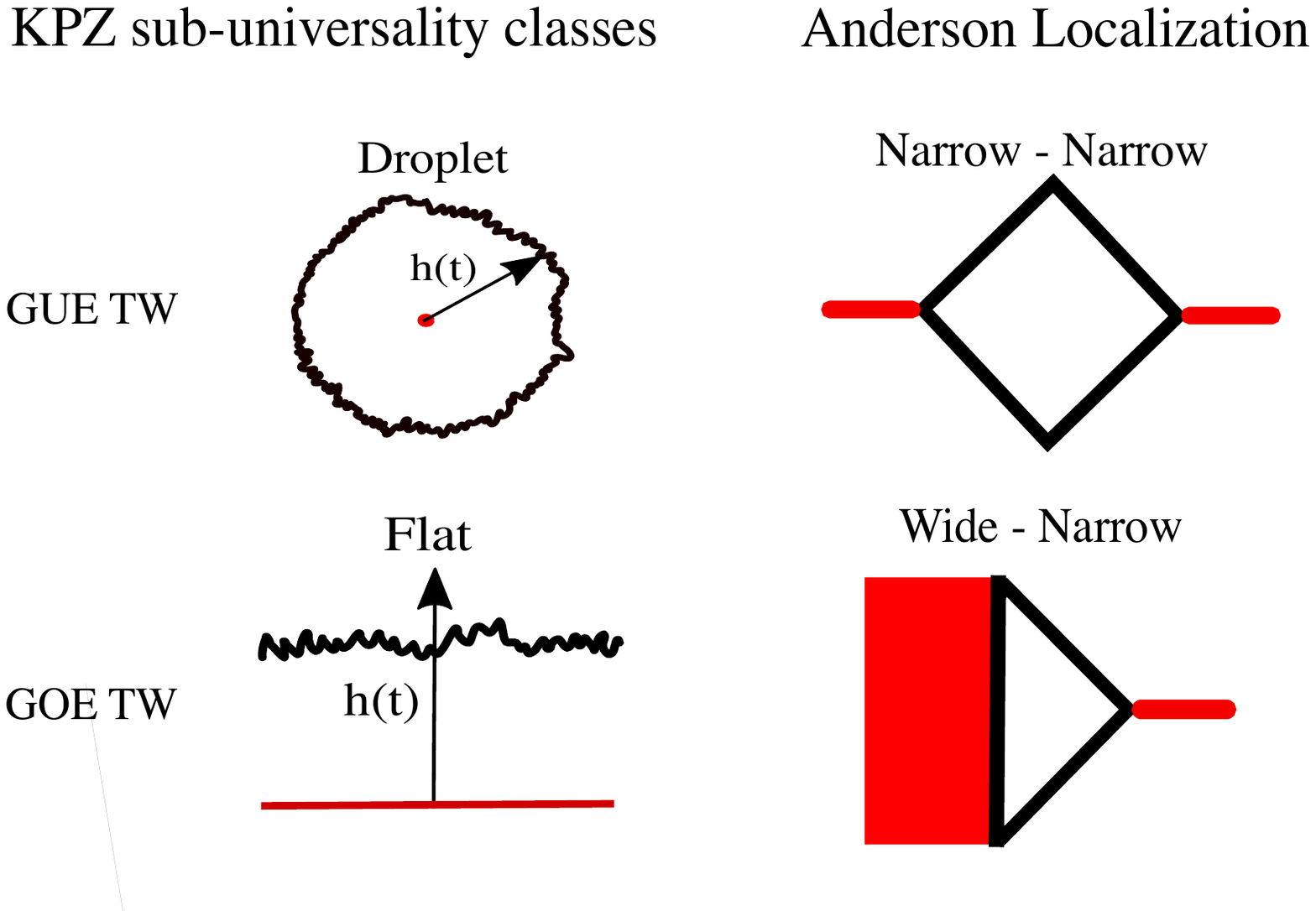}
\caption{Left: Illustration of the two KPZ sub-universality classes for rough interface growth fluctuations considered in this paper and their corresponding initial conditions: the GUE Tracy-Widom (TW) class with a droplet initial condition and the GOE TW class with a flat initial condition.
Right: Illustration of the two corresponding lead-scattering sample configurations for conductance fluctuations: `narrow-narrow' (NN) and `wide-narrow' (WN). The central region (square or rectangular) represents the scattering sample, while the left and right leads (shown in red) correspond to the input and output channels.
}
\label{fig1}
\end{center}
\end{figure}
%%%%%%%%%%%%%%%%%%%%%%%%%%%%%%%%%%%%%%%%%%%%%%%%%%%%%%%%%%%%%%%%%%%%%%%%%%%

In KPZ physics, the growth of a rough interface in 1+1 dimensions is described by the height function $h(x,t)$, where $x$ is the spatial coordinate and $t$ represents time \cite{KPZ_1986, HALPINHEALY1995, KPZ_30yrs_2015}, see Fig.~\ref{fig1}. The height function $h$ is a stochastic variable that evolves as $h(x,t) = v t + \alpha t^{1/3} \tilde{\chi}$, where $v$ is the growth velocity, $\alpha$ is a constant, and $\tilde{\chi}$ is a random variable of order one. The exponent $1/3$, which governs the algebraic growth of fluctuations, is a universal hallmark of KPZ physics, clearly observed in 2D Anderson localization, see \cite{PhysRevB.46.9984, prior2005conductance, Somoza_2007, prior2009conductance, Somoza_2015, Lemarie_PRL_2019, musen_2023} and this paper. More precisely, in 2D it is observed that the standard deviation \(\sigma\) of the conductance logarithm \(\ln g\) grows with system size \(L\) as \(\sigma \sim L^{1/3}\). 
This contrasts clearly with the behavior in (quasi-)1D Anderson localization, where the scaling exponent is \(1/2\), i.e., \(\sigma_{\text{1D}} \sim L^{1/2}\).

Central to this work is the property that the distribution of the fluctuating component $\tilde{\chi}$ belongs to different sub-universality classes depending on the initial condition of the interface \cite{KPZ_expt_Takeuchi_2010, KPZ_expt_Takeuchi2_2010, PhysRevLett.84.4882, Spohn2020kpz, Calabrese_2010, Dotsenko_2010, Prahofer_2000_PRL}. As illustrated in Figure \ref{fig1}, when starting from a flat initial condition ($h(x, t = 0) = 0$), the distribution of $\tilde{\chi}$ follows the Tracy-Widom (TW) distribution of the largest eigenvalue of random matrices in the Gaussian orthogonal ensemble (GOE). In contrast, if the interface begins as a droplet, the distribution aligns with the Tracy-Widom distribution of the Gaussian unitary ensemble (GUE). Other distributions are obtained for different initial conditions, which we will not discuss in this paper \cite{Baik_Rains_2000, PhysRevLett.108.190603, Takeuchi_2018}. The appearance of Tracy-Widom distributions in KPZ fluctuations is a significant and precise mathematical result in this framework \cite{Hartmann_2018_EPL}.

In this paper, we aim to definitively demonstrate the existence of sub-universality classes for conductance fluctuations. We show that these classes depend critically on the boundary conditions imposed on the 2D system yet are insensitive to the symmetry properties of the bulk. Changing the boundary conditions shifts the distribution of these fluctuations from the GOE to the GUE Tracy-Widom distributions. However, these conductance distributions remain unaffected by whether the localized system possesses time-reversal symmetry. Within the 10-fold way, breaking time-reversal changes the symmetry class from orthogonal to unitary.

At this stage, it is important to acknowledge the pioneering series of works~\cite{prior2005conductance, Somoza_2007, prior2009conductance, Somoza_2015}, which already addressed this problem. 
These studies demonstrated that the distribution of \(\ln g\) is non-Gaussian--unlike in (quasi-)1D Anderson localization--and is consistent with the GUE-TW distribution for a scattering configuration corresponding to the droplet KPZ case. 
They further argued that the distribution transitions to the Baik-Rains form in a different configuration, highlighting its sensitivity to boundary conditions.

Why, then, do we revisit this problem? 
Through a careful numerical analysis covering a broader range of system sizes and disorder strengths, we show that the scattering configuration and observable used to associate the conductance statistics with the Baik-Rains distribution—and, to a lesser extent, also with the GUE-TW distribution—are inadequate. 
As a result, definitive conclusions about the universality of conductance fluctuations in this regime, and their sensitivity to boundary conditions, cannot be drawn.

Furthermore, distinguishing the different universality classes of conductance fluctuations in the 2D Anderson localized regime is ultimately a question of measurement precision. The statistical accuracy achievable with standard sampling methods, as employed in previous studies, is insufficient to reach definitive conclusions. Given the fundamental importance of this question, we have developed a highly refined approach that enhances precision by up to 24 orders of magnitude. This unprecedented level of accuracy has enabled us to conclusively demonstrate that the universality class of localization fluctuations is determined by the boundary conditions rather than the symmetry class.

\begin{figure}
\begin{center}
\includegraphics[width=0.9\linewidth,angle=0]{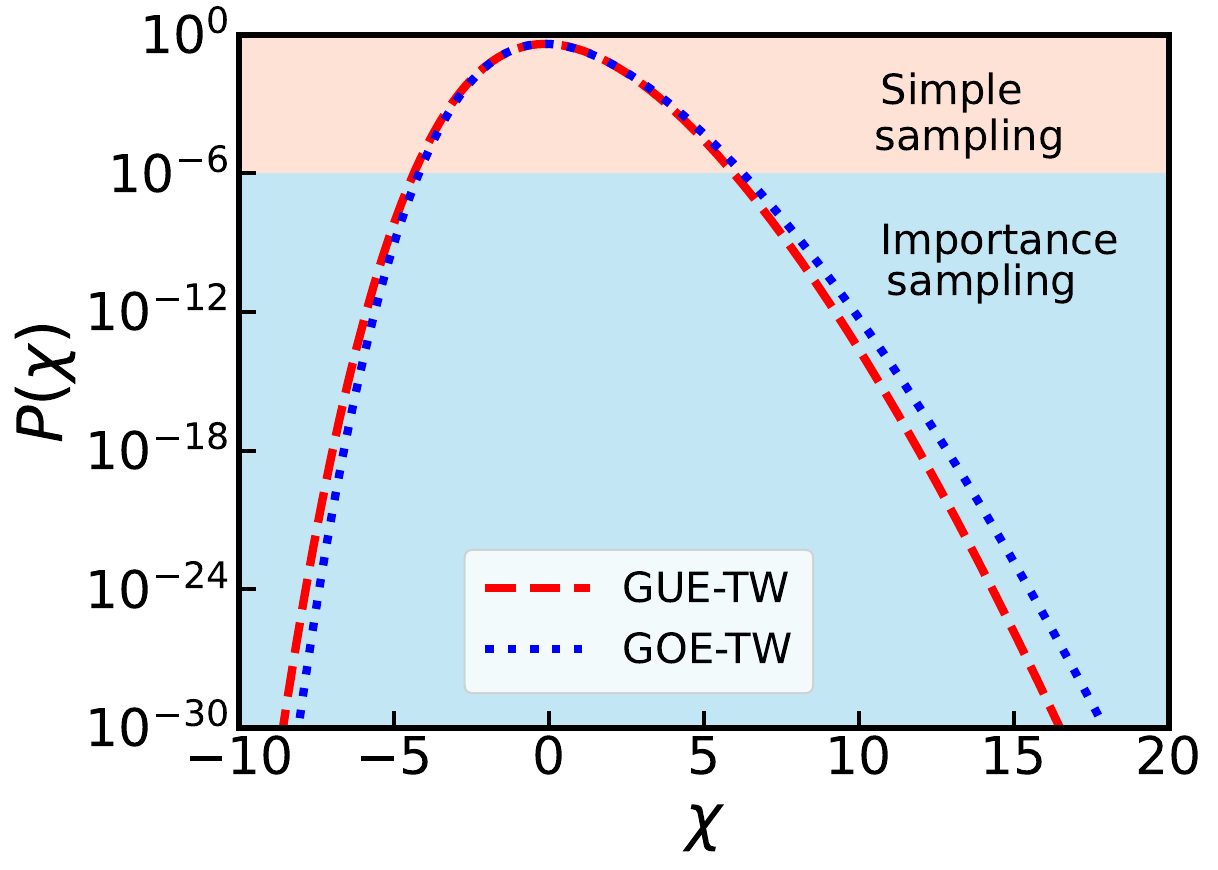}
\caption{GUE and GOE Tracy-Widom distributions~\cite{KPZ_data_Prahofer_Spohn}, rescaled such that their mean value is zero and their standard deviation is unity.
As clearly seen, these distributions are very similar, particularly in the regime accessible to simple sampling studies (typically $10^6$ disordered samples, leading to $P(\chi) \le 10^{-6}$ ), making them difficult to distinguish in numerical studies.
This is why we focus on the right tail region, which describes rare disorder configurations with anomalously large $\chi$ values, i.e., atypically large $\ln g$ values. 
The importance sampling scheme we implemented enables us to accurately describe the statistics of these rare events and distinguish between the two sub-universality classes.
}
\label{fig2}
\end{center}
\end{figure}

A major challenge arises due to the subtle differences between the GOE and GUE Tracy-Widom distributions. While these distributions have distinct means and standard deviations, when plotted for the rescaled variable $\chi \equiv (\tilde{\chi} - \langle \tilde{\chi} \rangle)/\sqrt{\langle \tilde{\chi}^2 \rangle-\langle \tilde{\chi }\rangle^2}$, with zero mean $\langle \chi \rangle=0$ and unit standard deviation $\sqrt{\langle \chi^2 \rangle}=1$, they appear nearly identical down to probabilities as low as $P(\chi) \ge 10^{-6}$, as shown in Figure \ref{fig2}. Notably, unlike in several studies about KPZ physics \cite{KPZ_expt_Takeuchi2_2010, takeuchi2012evidence, Somoza_2007, prior2009conductance}, we cannot distinguish sub-universality classes based on the mean or standard deviation, % of the fluctuating part of the (logarithm of the) conductance, which is analogous to the height $h(x,t)$ in Anderson localization,
as these values are non-universal and depend, for instance, on the strength of disorder.

This leads us to the objective of differentiating the shape of the distribution of the rescaled variable $\chi$ by determining $P(\chi)$ below the $10^{-6}$ threshold, ideally extending it to probabilities as low as $10^{-30}$. Achieving this level of precision is extremely challenging, as it requires sampling events that occur as rarely as once in $10^{30}$ instances. To accomplish this, one would typically need to calculate the conductance of $10^{30}$ 2D samples. Given the necessity of sufficiently large system sizes ($50 < L < 400$) to enter the strong localization regime, this goal is clearly unfeasible with standard numerical methods.

In this paper, we utilize a method known as importance sampling, specifically designed to address this challenge \cite{MC_book_Newman,MC_book_Landau}. The fundamental concept behind this approach involves biasing the selection of disorder configurations to focus on rare events that carry significance~\cite{Monthus_2006,Saito_2010,Hartmann_2018_EPL,Biroli_2022}.
Importantly, we have the means to unbias the final results.
By employing this technique, we can accurately describe the distributions of the logarithm of conductance, even for events as rare as $10^{-20}$, enabling us to clearly distinguish between the different distributions.

Following the analogy with KPZ physics, we investigate two distinct lead-system configurations, as illustrated in Fig.~\ref{fig1}. These configurations are anticipated, based on the analogy, to correspond to the different initial conditions (droplet and flat) in KPZ physics. We also explore the effects of a magnetic field, which breaks time-reversal symmetry.

%Our initial approach involves regular sampling of conductance fluctuations across different boundary and symmetry conditions, capturing $P(\chi) \ge 10^{-6}$. This initial sampling, however, is insufficient to decisively distinguish between the expected KPZ distributions. Therefore, we apply importance sampling through a Monte Carlo approach over disorder configurations for each case. The results from these simulations conclusively and affirmatively establish the Tracy-Widom GOE and GUE sub-universality classes of KPZ for conductance fluctuations in the 2D Anderson localized regime, dependent on boundary conditions yet unaffected by the symmetry class.

The remainder of the paper is organized as follows: In Section \ref{sec_model_method}, we detail our model and methodology, covering: \((i)\) the exact calculation of zero-temperature conductance in a 2D Anderson model, \((ii)\) the forward-scattering approximation and its suggested analogy between Anderson localization and the directed polymer problem, \((iii)\) a critical review of previous results on fluctuations in the Anderson localized regime and our rationale for considering them inconclusive, and \((iv)\) the importance sampling of disorder configurations necessary to achieve the precision required for a definitive conclusion on this problem. In Section \ref{sec_res}, we present the findings from both standard and importance sampling studies. Section \ref{sec:magfield} demonstrates the insensitivity of the conductance logarithm distribution to a magnetic field. Finally, we discuss the physical picture underlying our results, present our conclusions, and outline future research directions. The appendices provide further details on our approach and additional results.

%%%%%%%%%%%%%%%%%%%%%%%%%%%%%%%%%%%%%%%%%%%%%%%%%%%%%%%%%%%%%%%%%%%%%%%%%%%
%%%%%%%%%%%%%%%%%%%%%%%%%%%%%%%%%%%%%%%%%%%%%%%%%%%%%%%%%%%%%%%%%%%%%%%%%%%

\section{Model and methods}
\label{sec_model_method}

\subsection{Exact Calculation of Zero-Temperature Conductance in a 2D Anderson Model}\label{sec:model}

In this paper, we explore the 2D Anderson model on a square lattice defined by the Hamiltonian:
\begin{equation}
 H = \sum_i \varepsilon_i a_i^\dagger a_i 
   + \sum_{\langle i,j \rangle} (e^{i \Phi_{ij}} a_i^\dagger a_j + H.c.)\; .
\label{eq1}   
\end{equation}
Here \(a_i\), \(a_i^\dagger\) are the annihilation and creation operators of an electron at site \(i (x_i,y_i)\). The hopping amplitude is set to $1$, and the sum over $\langle i,j \rangle$ is restricted to nearest neighbors. 
 In the presence of a uniform magnetic field, Peierls phase factors are introduced
 \begin{equation}\label{eq:Peierls}
     e^{i \Phi_{ij}} = e^{[ i \Phi \frac{(x_i+x_j)(y_i - y_j)}{2}]}
 \end{equation}
where $\Phi$ is the magnetic flux per unit cell (see e.g. \cite{Nagaosa_PhysRevLett.70.1980,ohtsuki1993conductance,Ohtsuki_PhysRevB.51.10897,Furusaki_PhysRevLett.82.604}).

The site energies, \(\varepsilon_i \in [-W/2,W/2]\), are independent random variables with a uniform distribution. 
We also study Gaussian distributed disorder, where the site energies \(\varepsilon_i\) follow a normal distribution with zero mean and standard deviation $W$, described by
$
P(\varepsilon_i) \propto e^{-\frac{1}{2} \left(\frac{\varepsilon_i}{W}\right)^2}.
$
The results for Gaussian disorder are presented in Appendix~\ref{appendix_E}.

The key quantity of interest is the zero-temperature conductance of this scattering system. We consider a scattering system of size \(L \times L\) described by Eq.~\ref{eq1}, attached to two perfect (non-disordered) leads, one on the left boundary, the other on the right boundary, see Figs.~\ref{fig1} and \ref{app_fig_geometry}. Each lead can be either wide, a semi-infinite quasi-1D version of model Eq.~\ref{eq1} with \(\varepsilon_i =0\) for all sites \(i\), having the same section of the sample; or narrow, consisting of a 1D lead attached at the middle of one edge \cite{Somoza_2007,Lemarie_PRL_2019}.

We compute the zero-temperature, dimensionless conductance at energy $E_F$ exactly using the Fisher–Lee formula \cite{fisher1981relation}:
\begin{equation}
    g(E_F) = \text{Tr}[\Gamma_L \mathcal G^r \Gamma_R \mathcal G^a ] 
\end{equation}
which uses the Green's function \(\mathcal{G}(E_F)\) of the scatterer dressed by the leads, with the exponent \(r\) (\(a\)) denoting retarded (advanced) Green's function. \(\Gamma_{L,R}=-2 \text{Im}[\Sigma_{L,R}^r]\) is the imaginary part of the self-energies associated with the leads, see \cite{datta1997electronic}. The Green's function is efficiently calculated numerically using the recursive Green's function approach \cite{datta1997electronic}. Some of our calculations were benchmarked using the Kwant library \cite{Kwant_2014}, but the results presented in this paper were obtained using a custom code we developed.

\subsection{Forward-Scattering Approximation and Analogy with the Directed Polymer Problem} \label{sec:FSA}

The Green's function element at energy \(E_F\) between a point \(a\) on the left edge of the scattering system and a point \(b\) on the right edge can be approximated using the locator expansion \cite{Anderson:PR58}:
\begin{equation}
  \mathcal G_{a,b}(E_F) \approx \sum_{\mathcal P} \prod_{i\in\mathcal P} (E_F-\varepsilon_i)^{-1}  
\end{equation}
where the sum is over all paths $\mathcal P$ connecting $a$ to $b$, and the product is over all sites belonging to path $\mathcal P$. In the limit of strong disorder $W/t \gg 1$, the weight $\prod_{i\in\mathcal P} (E_F-\varepsilon_i)^{-1}$ of a path $\mathcal P$ will decrease exponentially with its length. The sum over paths will then be dominated by the forward-scattering paths that propagate from left to right, i.e., in the direction from $a$ to $b$ \cite{nguyen1985tunnel, fritzsche1990hopping, medina1989interference, prior2005conductance, Somoza_2007, prior2009conductance, PhysRevB.93.054201}. We can then further approximate the Green's function $\mathcal G_{a,b}(E_F)$ as:
\begin{equation}\label{eq:FSA}
  \mathcal G_{a,b}(E_F) \approx \sum_{\mathcal DP} \prod_{i\in\mathcal DP} (E_F-\varepsilon_i)^{-1}  
\end{equation}
which is known as the Nguyen-Spivak-Shklovskii (NSS) model \cite{nguyen1985tunnel}.

While we discuss the forward-scattering approximation to provide an intuitive connection to the directed polymer problem, it is important to note that this approximation neglects crucial loop interference effects and incorrectly predicts an Anderson transition in one and two dimensions. It is therefore uncontrolled beyond the regime of strong disorder \cite{PhysRevB.93.054201}. In contrast, our numerical work is exact and fully accounts for all loop interference effects. As we will show, it nevertheless confirms the analogy with the directed polymer problem and KPZ physics that the forward-scattering approximation suggests is valid.

The expression \eqref{eq:FSA} for the Green's function closely resembles that of the partition function of a directed polymer in a random potential \cite{HALPINHEALY1995, PhysRevB.43.10728}:  
\begin{equation}    
Z = \sum_{\mathcal {DP}} \prod_{i\in\mathcal {DP}} e^{-\beta V_i}  \;,
\end{equation}
where \(\beta\) is the inverse temperature, and \(V_i\) represents the on-site random energies. This correspondence is established for \(\beta=1\) and \(V_i = \ln (E_F-\varepsilon_i)\). Notably, we consider \(E_F = 0\), allowing \((E_F-\varepsilon_i)\) to be negative, which results in complex on-site energies for the directed polymer \cite{PhysRevLett.62.979, Zhang_1989, medina1989interference, GELFAND199167, gangopadhyay2013magnetoresistance}.  

The directed polymer (DP) problem in dimension \(1+1\) is a well-known classical statistical mechanics problem that corresponds exactly to KPZ physics \cite{Kardar_Zhang_1987,derrida1988polymers,halpin1995kinetic}. In particular, the free energy of the DP is directly related to the height of the interface in the KPZ problem. Numerical simulations suggest that directed polymers with complex on-site random energies also belong to the KPZ universality class \cite{PhysRevB.46.9984, PhysRevLett.62.979, Zhang_1989, medina1989interference, GELFAND199167, gangopadhyay2013magnetoresistance}.  

One can study this DP problem on a 2D square lattice. Typically, this is done by considering the propagation of the DP along one diagonal, chosen as the time direction, with the transverse direction (along the other diagonal) corresponding to the spatial coordinate \cite{PhysRevA.44.R3415}. This choice of propagation along the diagonal is not arbitrary: in this configuration, all paths from one corner to the opposite corner of a square sample, or from one diagonal to the opposite corner, have the same length. As we shall see, these configurations are also highly relevant for conductance fluctuations.  

The partition function depends on space \(x\) and time \(t\), i.e., \(Z(x,t)\), and also on the chosen initial condition at \(t=0\).  
By analogy with the forward-scattering approximation for the Green's function, the time direction of the directed polymer corresponds to the longitudinal direction between the two leads (\(t \equiv L\), with $L$ the (network) distance between the leads, see Fig.~\ref{app_fig_geometry}), while the spatial direction of the polymer corresponds to the transverse direction. In our setup, we consider a narrow lead at the right corner, which corresponds to evaluating \(Z(x=0,t=L)\) in the directed polymer framework.  

Consequently, the forward-scattering approximation suggests that the logarithm of the conductance, \(\ln g\), behaves analogously to the height \(h\) of the KPZ interface or, equivalently, the free energy \(F=-\ln Z\) (with unit temperature) of the directed polymer problem. This leads to the following relation \cite{PhysRevB.46.9984, prior2005conductance, Somoza_2007, prior2009conductance, Somoza_2015, Lemarie_PRL_2019, musen_2023, Monthus_2012}:  
\begin{equation}\label{eq:lngvsL}
    \ln g=-2 L/\xi + \alpha\left(L/\xi\right)^{1/3} \tilde{\chi} \;,
\end{equation}
where \(\xi\) is the localization length, and \(\tilde{\chi}\) is a random variable of order one. The first term on the right-hand side describes the exponential decay of the conductance with \(L\), a hallmark of Anderson localization, which corresponds to the linear growth of a KPZ interface with velocity \(v \propto 1/\xi\). The second term captures the nontrivial growth of fluctuations, scaling algebraically with \(L\) (or time in the KPZ/DP picture) and following the universal KPZ exponent of \(1/3\).

Moreover, the distribution of the rescaled variable $\tilde{\chi}$ is expected to be universal, with a universality class dependent on the boundary/initial condition. In the DP problem, one can start from a single point or a line, corresponding to the droplet or flat initial conditions of KPZ, and to a narrow left lead or a wide left lead in the conductance problem. It was shown for the KPZ and DP problems that, in the droplet case, $\tilde{\chi}$ follows the Tracy-Widom (TW) distribution of the largest eigenvalue of random matrices in the Gaussian unitary ensemble (GUE), while it follows the GOE TW distribution in the flat case \cite{Johansson2000, Prahofer_2000_PRL, KPZ_expt_Takeuchi_2010, KPZ_expt_Takeuchi2_2010, takeuchi2012evidence, Sasamoto_1D_KPZ_rev_2016, Hartmann_2018_EPL, Takeuchi_2018}. This is illustrated in Fig.~\ref{fig1}.

\subsection{Critical Review of Previous Results on Universal Fluctuations of the Conductance Logarithm in the 2D Anderson Localized Regime}

Several predictions based on the analogy with DP/KPZ physics have been confirmed numerically for 2D Anderson localization \cite{PhysRevB.46.9984, prior2005conductance, Somoza_2007, prior2009conductance, Somoza_2015, Lemarie_PRL_2019, musen_2023}. In particular, these include the non-trivial scaling of the standard deviation of the logarithm of the conductance:  
\begin{equation}
    \sigma \equiv \sqrt{\langle (\ln g)^2\rangle -\langle \ln g\rangle^2 } \sim L^{1/3} \;,
\end{equation}  
and the scaling of the disorder average of \(\ln g\):  
\begin{equation}
    \langle \ln g \rangle \approx -\frac{2L}{\xi} + \alpha \left(\frac{L}{\xi}\right)^{1/3} \langle \tilde{\chi} \rangle \;,
\end{equation}  
see Appendix \ref{sec.appFSS} and Fig. \ref{fig_app_scaling}.

The seminal numerical studies \cite{prior2005conductance, Somoza_2007, prior2009conductance} used different lead configurations compared to the present study: (i) a narrow-narrow configuration denoted \(\tilde{N}\tilde{N}\), which involves attaching 1D leads to the middle of the opposite sides of the square sample (instead of the opposite corners as in Fig.~\ref{fig1}); (ii) a wide-wide configuration denoted \(\tilde{W}\tilde{W}\), where the leads span the same width as the square sample and are attached to two opposite sides (with periodic boundary conditions perpendicular to the leads). 

In the \(\tilde{N}\tilde{N}\) configuration, \(\langle \ln g \rangle\) exhibits an \(L^{1/3}\) correction to the linear \(L\) decay. Indeed, the distribution of the fluctuating part \(\tilde{\chi}\) is consistent with the GUE-TW distribution, which has a nonzero mean, \(\langle \tilde{\chi} \rangle \neq 0\).  

In contrast, for the \(\tilde{W}\tilde{W}\) configuration, no \(L^{1/3}\) correction to \(\langle \ln g \rangle\) is observed. Indeed, the data align with the Baik-Rains distribution, which in KPZ physics corresponds to the stationary case. This distribution has a zero mean, \(\langle \tilde{\chi} \rangle = 0\).

However, the result of a Baik-Rains distribution for the \(\tilde{W}\tilde{W}\) configuration is surprising. In KPZ physics, the stationary case corresponds to a specific initial condition \cite{PhysRevLett.108.190603, borodin2015height, PhysRevLett.124.250602}, namely a two-sided Brownian motion in the spatial direction \(x\). This state is achieved when the correlation length of the KPZ spatial fluctuations exceeds the system size. This KPZ stationary distribution seems quite distinct from that of a wide lead, which corresponds instead to a flat initial condition, characterized by the absence of fluctuations at \(L \equiv t = 0\). For Anderson localization, we expect that a KPZ stationary distribution would correspond to a quasi-1D regime with the localization length being much smaller than the transverse size—a regime that, to the best of our knowledge, has not been numerically studied due to its computational difficulty. Finally, we note that in KPZ physics, the Baik-Rains distribution describes the height difference \(h(x,t) - h(x,0)\), rather than the height \(h(x,t)\) itself.  

Another intriguing issue with the $\tilde{N}\tilde{N}$ configuration studied in \cite{prior2005conductance, Somoza_2007, prior2009conductance} is that at strong disorder, when the localization length $\xi$ becomes comparable to the lattice spacing, the network's structure plays a significant role. Since KPZ/DP physics arises from the optimization of the disorder cost across different paths, a key requirement is that all paths must have the same length. Otherwise, shorter paths could be exponentially favored, distorting the fluctuations. In the $\tilde{N}\tilde{N}$ configuration, however, the paths do not have uniform lengths: the straight path along the direction of the leads is clearly the shortest.

Due to these concerns, we have also simulated the \(\tilde{N}\tilde{N}\) and \(\tilde{W}\tilde{W}\) configurations, extending the analysis to a broader range of parameters. Our results are presented in Appendix~\ref{sec.appSOP}, Fig.~\ref{fig:SOPTW}. When reproducing the same parameters used in previous studies \cite{prior2005conductance, Somoza_2007, prior2009conductance}, we recover similar results. However, by exploring a wider range of system sizes \(L\) (up to 400), disorder strengths \(W\) (from 10 to 50), and disorder types (uniform and Gaussian), we observe systematic deviations from the expected GUE-TW and Baik-Rains distributions.

These deviations are most clearly captured by analyzing the skewness of the distributions---a measure of their asymmetry around the mean---as shown in Fig.~\ref{fig:SOPTW}. For the \(\tilde{W}\tilde{W}\) configuration, such deviations persist across all disorder strengths in the large-size limit. In contrast, for the \(\tilde{N}\tilde{N}\) case, deviations appear only at strong disorder, where the localization length \(\xi(W) \leq 1\). Crucially, the deviations depend on the type of disorder distribution, even at large disorder values where KPZ physics is expected to hold---questioning the universality of the fluctuations. 

We interpret these deviations as evidence that the \(\tilde{W}\tilde{W}\) configuration does not correspond to the Baik-Rains distribution. The \(\tilde{N}\tilde{N}\) configuration matches the GUE-TW distribution only at moderate disorder strengths, where the structure of the network is not expected to play a significant role. Hence, these configurations do not allow us to determine whether conductance fluctuations are truly universal in the 2D localized regime, and whether they depend sensitively on boundary conditions, as suggested by the analogy with KPZ physics.

Instead, we investigate the \(NN\) and \(WN\) configurations and show that they yield universal distributions: GUE-TW and GOE-TW, respectively. The \(NN\) configuration differs from the \(\tilde{N}\tilde{N}\) setup in that the leads are attached at opposite corners rather than at the midpoints of opposite sides, ensuring that all paths from the left to right boundaries have the same length. This property is crucial for studying conductance fluctuations in the localized regime, as shorter paths contribute exponentially more than longer ones. Indeed, the universal fluctuations predicted in KPZ physics arise from an optimization over different paths having the same length, but different energies controlled by the disorder. 
Similarly, the \(WN\) configuration shares this essential feature.

%%%%%%%%%%%%%%%%%%%%%%%%%%%%%%%%%%%%%%%%%%%%%%%%%%%%%%%%%%%%%%%%%%%%%%%%%%%
\begin{figure*}[t]
\begin{center}
\includegraphics[width=0.8\linewidth,angle=0]{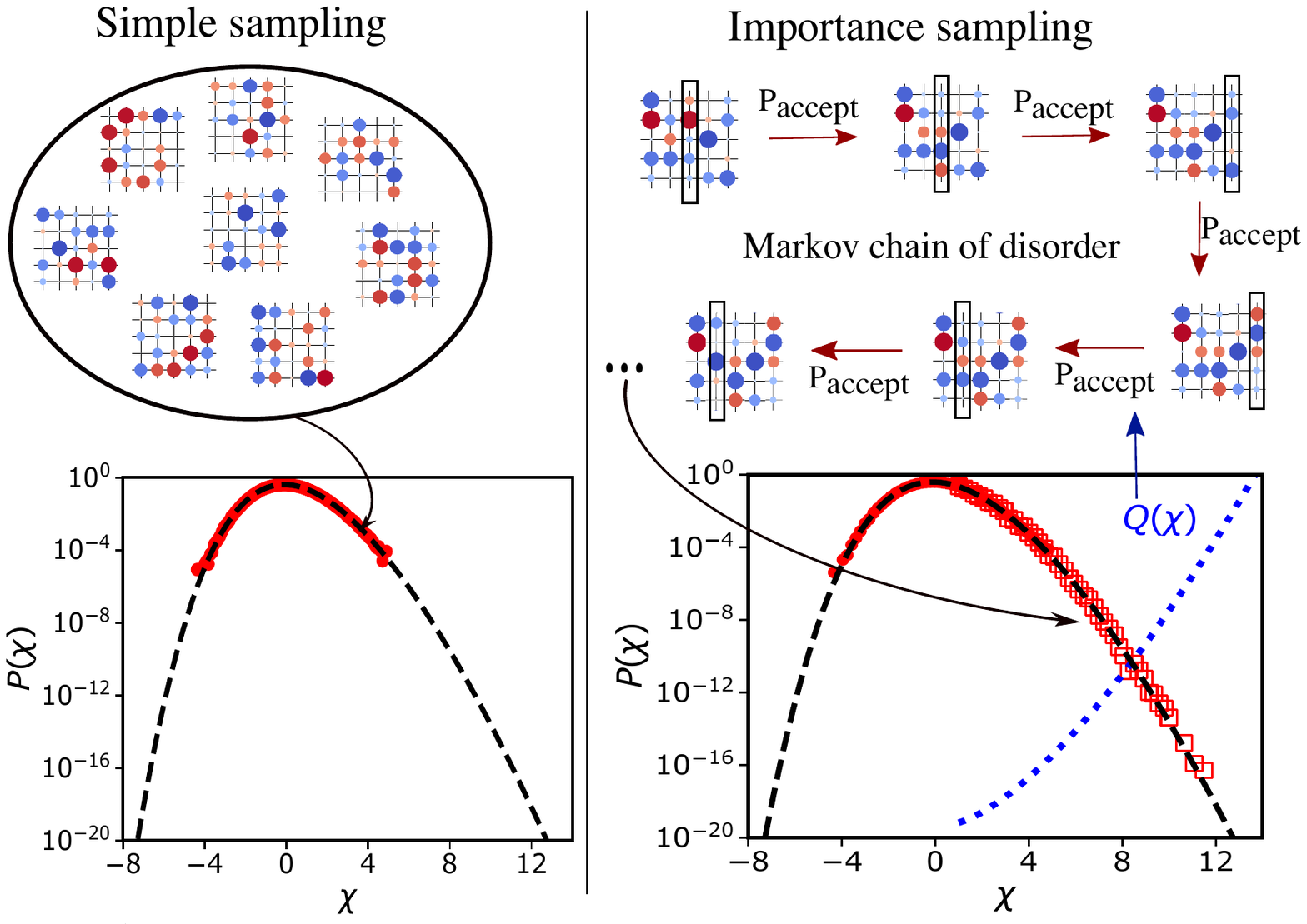}
\caption{Schematic comparison of simple sampling and importance sampling. 
With \(10^6\) independent disorder configurations, simple sampling is limited to resolving probabilities down to \(P(\chi) \approx 10^{-6}\), making rarer events inaccessible. In contrast, importance sampling can capture extremely rare fluctuations, reaching probabilities as low as \(P(\chi) \sim 10^{-30}\) in this work.
This is achieved using a Markov chain Monte Carlo approach: starting from a random disorder realization, disorder values are updated on randomly selected columns following the Metropolis acceptance criterion, which depends on a weight function \(Q(\chi)\). After approximately \(10^6\) update attempts, rare events are efficiently sampled, extending the accessible probability range by many orders of magnitude.
}
\label{fig:illustration}
\end{center}
\end{figure*}
%%%%%%%%%%%%%%%%%%%%%%%%%%%%%%%%%%%%%%%%%%%%%%%%%%%%%%%%%%%%%%%%%%%%%%%%%%%

%%%%%%%%%%%%%%%%%%%%%%%%%%%%%%%%%%%%%%%%%%%%%%%%%%%%%%%%%%%%%%%%%%%%%%%%%%%
\begin{figure*}
\begin{center}
\includegraphics[height=4.5cm,width=15cm,angle=0]{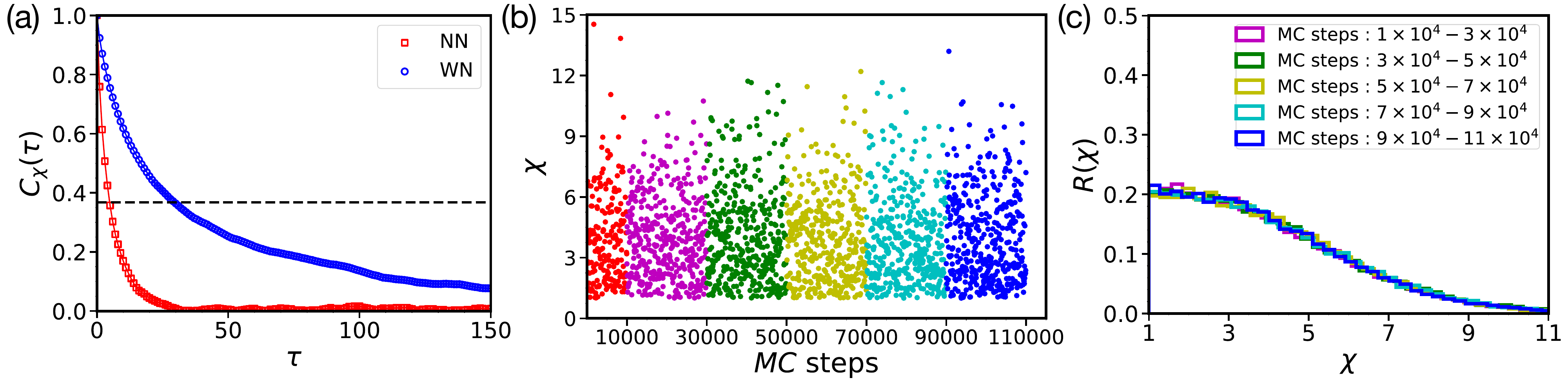}
\caption{(a)~Auto-correlation function of an importance sampling Markov chain of \(\chi\) with Monte Carlo time separation \(\tau\). We consider the `narrow-narrow' system with size \(L=202\) (NN, in red) and `wide-narrow' system with \(L=101\) (WN, in blue) lead geometries (see Fig.~\ref{app_fig_geometry}), and a uniform disorder distribution of strength \(W=50\). The auto-correlation function demonstrates that for large Monte Carlo time separations, the \(\chi\) values become essentially uncorrelated/independent. The dashed line indicates the threshold \(C_{\tau}(\chi) = 1/e\), used to estimate the auto-correlation time. After equilibration (see (b)), we will use the \(\chi\) values every 50 and 100 Monte Carlo steps for the NN and WN lead geometries, respectively, both of which are significantly larger than the auto-correlation time, for the measurement process.
(b)~We show the resulting data for the uncorrelated \(\chi\) values (i.e., sampled every 50 Monte Carlo steps) for the NN lead geometry (with the same parameters as in (a)), during equilibration of \(10^4\) Monte Carlo steps (red dots), and subsequent measurement windows of duration \(2 \times 10^4\) Monte Carlo steps. The occurrence of disorder configurations with larger \(\chi\) values and stationary fluctuations is clearly observed after equilibration.
(c)~Histograms \(R(\chi)\) corresponding to the different measurement windows shown in (b) all collapse onto a single distribution, demonstrating that the Markov chain has reached a steady state. The data are for the NN lead geometry with the same parameters as in (a).
}
\label{fig:datacorr}
\end{center}
\end{figure*}
%%%%%%%%%%%%%%%%%%%%%%%%%%%%%%%%%%%%%%%%%%%%%%%%%%%%%%%%%%%%%%%%%%%%%%%%%%%

\subsection{Importance sampling of disorder configurations} \label{sec:impsamp}

The primary objective of this study is to precisely determine the distribution of the logarithm of conductance, $\ln g$, and to elucidate its fluctuations under different disorder configurations. As mentioned earlier, our goal is to ascertain whether the distinct cases of `narrow-narrow' and `wide-narrow' correspond to different KPZ sub-universality classes, specifically the GUE and GOE Tracy-Widom distributions, respectively. Since the localization length, $\xi$, and the constant, $\alpha$, in Eq.~\eqref{eq:lngvsL} are non-universal (dependent on the disorder strength or distribution), we find it necessary to work with the rescaled conductance logarithm,
\begin{equation} \label{eq:chi}
\chi \equiv \frac{\ln g - \langle \ln g \rangle}{\sigma}\;, 
\end{equation}
which has zero mean and unit standard deviation.
Given the extreme similarity between the GOE and GUE TW distributions for $\chi$ (see Fig.\ref{fig2}), differentiating between them necessitates sampling exceedingly rare events, occurring as infrequently as $10^{-30}$.

{The typical method for determining the distribution of $\chi$ involves randomly sampling disorder configurations, see Fig.~\ref{fig:illustration} left panels. This process includes drawing $N$ independent disordered samples $D_1$, $D_2$, ..., $D_N$, and computing their corresponding $\chi$ values, $ \chi_1$, $ \chi_2$, ..., $ \chi_N$. However, the histogram obtained through this method cannot effectively characterize the distribution's tails for frequencies smaller than $1/N$ due to the rarity of these events. Addressing frequencies as low as $10^{-30}$ would require an impractical $10^{30}$ samples using this simple sampling.}

A well-established approach to addressing this challenge is importance sampling \cite{Monthus_2006,Hartmann_2018_EPL,Hartmann_2020_PRE,Biroli_2022}, illustrated in the right panels of Fig.~\ref{fig:illustration}. In our context, this method strategically focuses on rare, significant disorder configurations by allocating more samples to relevant events, allowing for an accurate reconstruction of their true distribution. In this section, we detail the importance sampling approach used in this paper.

As illustrated in Fig.~\ref{fig:illustration}, the importance sampling approach in our study revolves around two fundamental concepts: (i) Instead of examining independent disorder configurations, we employ a Markov chain within the space of disorder configurations (Metropolis–Hastings algorithm)~\cite{Metropolis_2004}. This involves transitioning from one disorder configuration to another using a specific acceptance probability. (ii) Crucially, the acceptance probability is flexible and can be tailored to our requirements. Given our focus on rare events, we choose an acceptance probability that favors disorder configurations corresponding to the tails of the distribution, ensuring a targeted exploration of these significant occurrences. 

Starting from an initial disorder configuration $D_{i}$, we create a trial disorder configuration $D_{t}$ by randomly selecting a column of onsite energies (a total of $L_y$ values) and replacing them with new values randomly drawn from the original disorder distribution. We calculate $\chi$ both before ($\chi(D_i)$) and after ($\chi(D_t)$) changing the column of disorder.

The decision to accept or reject the change from the initial configuration $D_{i}$ to the trial configuration $D_{t}$ is made based on the acceptance probability:
\begin{equation}
    P_{\rm accept}(D_t | D_i) = \min\left [\frac{Q(\chi(D_t))}{Q(\chi(D_i))},1\right ]
\end{equation}
Here, the function $Q$ is a user-defined function chosen to serve our specific purpose. This Markov chain will eventually converge to a stationary distribution, where a disorder configuration $D$ is visited with a probability proportional to $Q(\chi(D))$. Thus the histogram $R(\chi)$ of the $\chi$ values sampled by the stationary distribution of the Markov process is $R(\chi) \propto P(\chi) Q(\chi)$, where $P(\chi)$ is the actual distribution of $\chi$ when the disorder configurations are chosen independently, see Fig.~\ref{fig:datacorr}. 
Consequently, we can obtain the actual distribution $P(\chi)$ from the histogram $R(\chi)$ as $P(\chi) \propto R(\chi)/Q(\chi)$ up to a normalization factor.

We aim to emphasize rare events in the tails of \( P(\chi) \) by appropriately choosing \( Q(\chi) \).  
If we set \( Q(\chi) \) as the inverse of the actual distribution of \( \chi \), i.e.,  
\( Q(\chi) = 1/P(\chi) \), each \( \chi \) is sampled with equal probability, resulting in a flat histogram \( R(\chi) \). This choice is the most effective.  
In practice, since \( P(\chi) \) is unknown, a suitable \( Q(\chi) \) can be estimated using the simple sampling distribution. However, it is important to note that the resulting distribution \( P(\chi) \) should remain independent of the specific choice of \( Q(\chi) \); only the range of accessible \( \chi \) values is affected. We systematically verified this property by exploring various, less optimal, choices of \( Q(\chi) \), as shown in Fig.~\ref{fig_appD}.

The simple sampling of conductance fluctuations is a crucial first step in the importance sampling procedure. Firstly, this is because the mean ($\langle \ln g \rangle$) and variance ($\sigma^2 = \langle \ln g^2 \rangle - \langle \ln g \rangle^2$) of $\ln g$ are necessary to compute the rescaled conductance logarithm (see Eq.\eqref{eq:chi}), and they can be accurately obtained through simple sampling. Secondly, the simple sampling distribution is instrumental in determining an appropriate guiding function $Q(\chi)$. The approach we use follows Ref.~\cite{Monthus_2006} and involves fitting the simple sampling distribution in the right tail ($\chi > 0 $) using the function 
\begin{equation}\label{eq:guiding}
    1/Q(\chi) \equiv e^{a -b|\chi|^{\eta} + c \ln|\chi|} \; ,
\end{equation}
where $\eta= -3/2$ and $a$, $b$ and $c$ are adjustable parameters. 
Finally, the results obtained from simple sampling enable us to determine the normalization factor of the importance sampling distribution and validate its accuracy near the center of the distribution where simple sampling results are sufficiently precise.

The Monte Carlo process described above constitutes a Markov chain in the disorder configuration space, where a single Monte Carlo (MC) sweep involves updating $L$ columns of disorder, ensuring that the disorder at each site has been updated. Once a sufficiently large number of MC sweeps (typically $10000$ MC sweeps) is performed, this Markov process has converged to a stationary state.
However, because the successive disorder configurations encountered in our Monte Carlo scheme are not necessarily independent, we calculate the autocorrelation time, see Fig.~\ref{fig:datacorr}. Fluctuations in the system are considered meaningful only after a number of time steps greater than the autocorrelation time, ensuring that we are analyzing statistically uncorrelated data points.

The importance sampling process allows us to work on any fixed regime of conductance fluctuations $\chi \in [\chi_{min}, \chi_{max}]$ where we can freely choose $\chi_{min}$, or $\chi_{max}$ without modifying our scheme. For our calculations we focus on the positive tail of the conductance fluctuations ($\chi_{min} \geq 1$).

\section{Results}
\label{sec_res}

In this section, we report the statistics of conductance fluctuations, first in the orthogonal symmetry class with $\Phi=0$ and then in the unitary symmetry class with $\Phi\ne 0$ (see Eqs.~\eqref{eq1} and \eqref{eq:Peierls}), considering the two different lead configurations $-$ narrow-narrow (NN) and wide-narrow (WN) shown in Fig.~\ref{fig1}. Our focus is on the strongly localized regime with $\xi \ll L$, where the analogy with KPZ physics is expected to hold~\cite{PhysRevB.46.9984, prior2005conductance, Somoza_2007, prior2009conductance, Somoza_2015, Lemarie_PRL_2019, musen_2023}. We consider large disorder strengths up to $W = 50$, with varying system sizes up to $L\approx 400$.

\subsection{Simple sampling}

%%%%%%%%%%%%%%%%%%%%%%%%%%%%%%%%%%%%%%%%%%%%%%%%%%%%%%%%%%%%%%%%%%%%%%%%%%%
\begin{figure}[t]
\begin{center}
\includegraphics[width=\linewidth,angle=0]{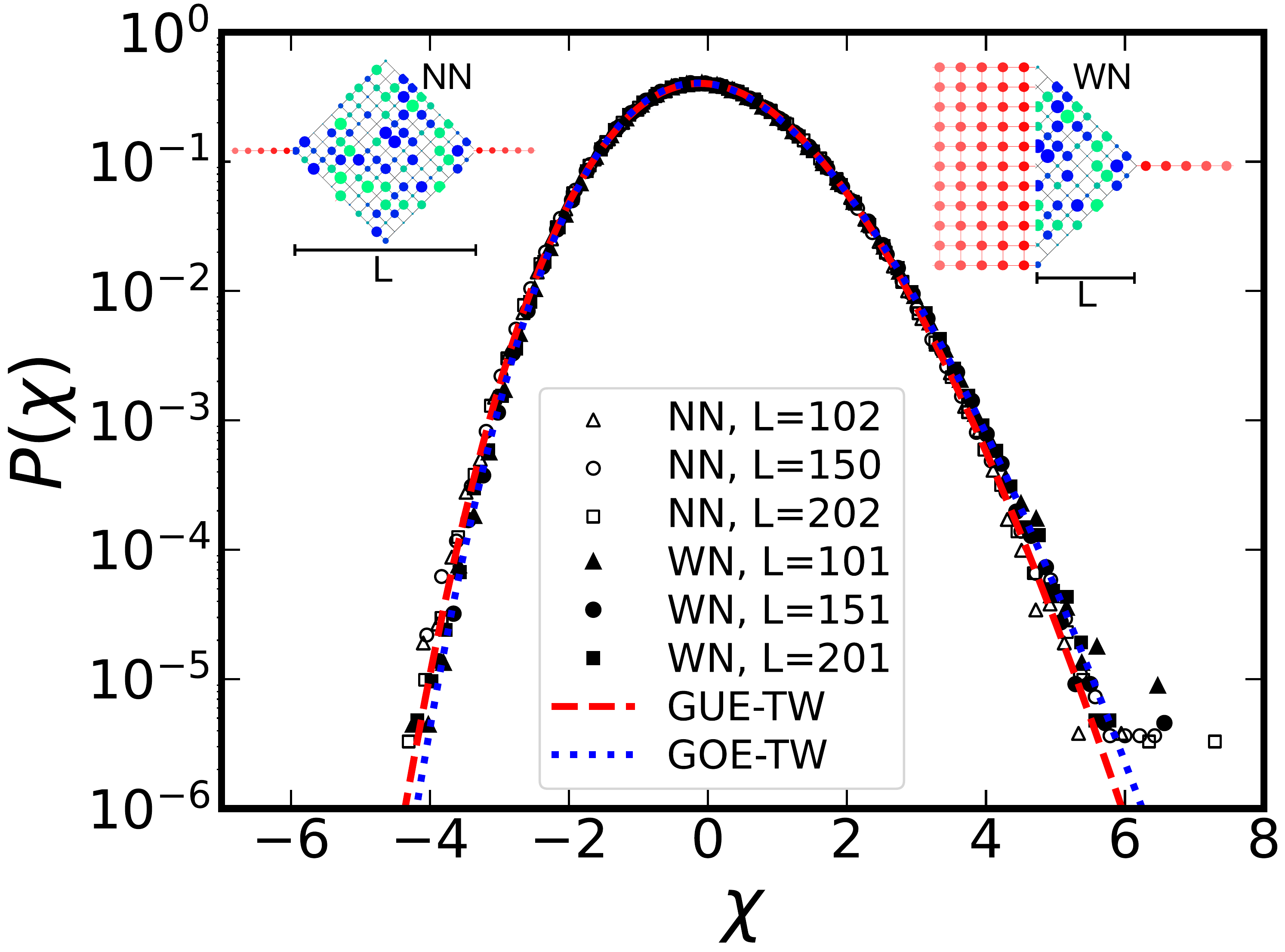}
\caption{Simple sampling results:
The distribution of the rescaled conductance logarithm 
\(\chi = (\ln g - \langle \ln g \rangle)/\sigma\)
is shown for the narrow-narrow (NN, open symbols) and wide-narrow (WN, solid symbols) lead geometries. 
A total of \(10^6\) independent disorder realizations were generated from a uniform distribution with a disorder strength of \(W = 50\).
The scattering region and the attached leads at both ends are shown in the inset for clarity: For the NN case, the scattering region corresponds to the 2D Anderson model on a rotated square lattice of size \(L/2 \times L/2\), with \(L = 102, 150, 202\). 
For the WN case, the scattering region corresponds to the 2D Anderson model on the right half of a rotated square lattice of size \(L \times L\), with \(L = 101, 151, 201\); see also Fig.~\ref{app_fig_geometry}. 
We observe that the data (NN, open symbols; WN, solid symbols) obtained via the simple sampling procedure align nicely with the GUE (long-dashed red line) and GOE (blue dotted line) Tracy-Widom distributions. However, within the range where \(P(\chi) > 10^{-6}\), it is impossible to distinguish clearly between the numerical distributions for these two different cases NN and WN.
}
\label{fig:GOUETWsimple}
\end{center}
\end{figure}
%%%%%%%%%%%%%%%%%%%%%%%%%%%%%%%%%%%%%%%%%%%%%%%%%%%%%%%%%%%%%%%%%%%%%%%%%%%

%%%%%%%%%%%%%%%%%%%%%%%%%%%%%%%%%%%%%%%%%%%%%%%%%%%%%%%%%%%%%%%%%%%%%%%%%%%
\begin{figure}
\centerline{
\includegraphics[height=6.0cm,width=8.0cm,angle=0]{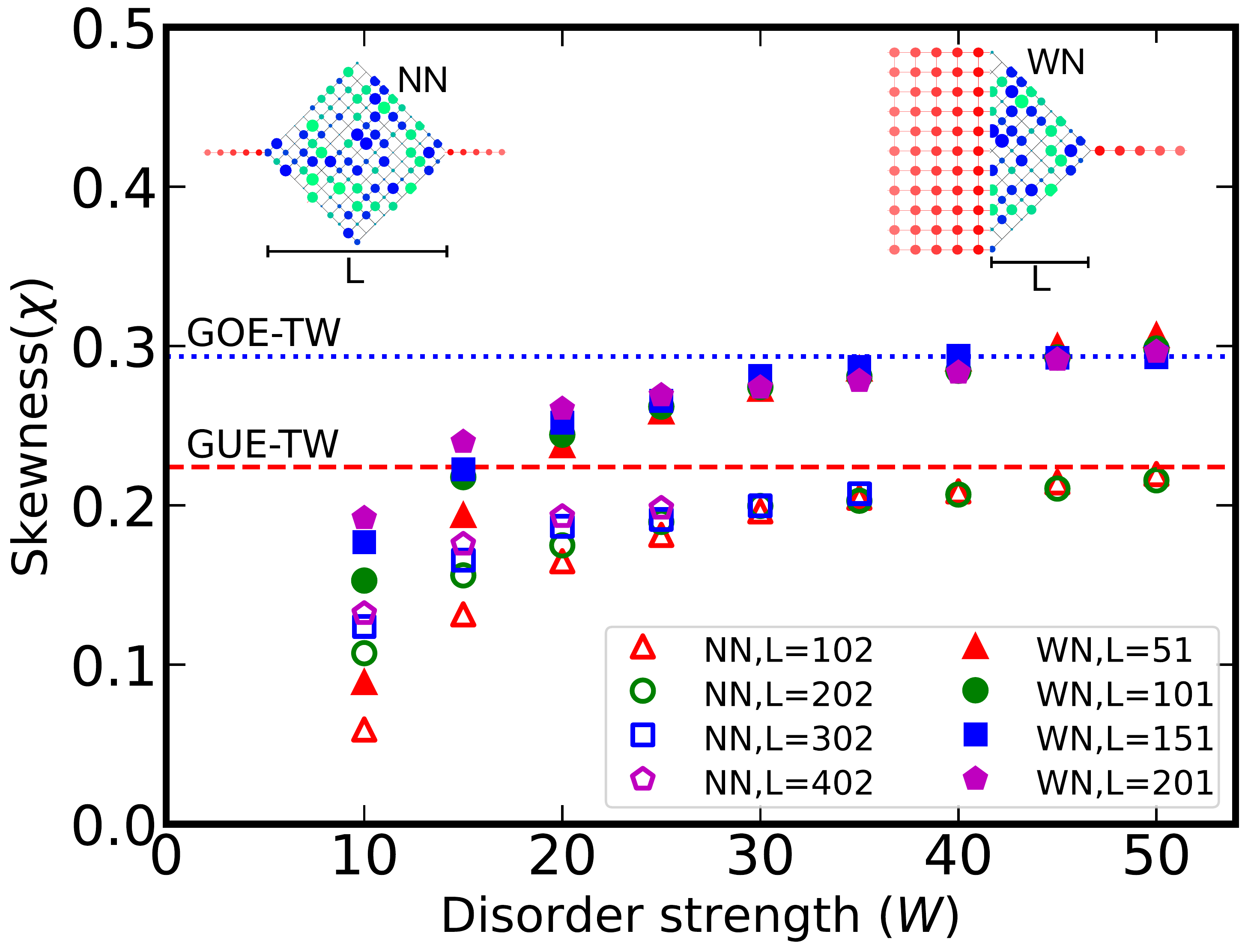}
}
\caption{
Variation of the skewness of the distribution of the rescaled conductance logarithm \( P(\chi) \) with disorder strength and system size for the narrow-narrow (NN) and wide-narrow (WN) lead geometries, as illustrated in the insets. The skewness is computed from a simple sampling of \( 10^6 \) independent disorder realizations.  
The dashed red line and dotted blue line represent the skewness values of the GUE and GOE Tracy-Widom (TW) distributions, respectively.  
The computed skewness values converge to these expected TW values in the limits of large disorder and large system size.  
}
\label{fig_skew}
\end{figure}
%%%%%%%%%%%%%%%%%%%%%%%%%%%%%%%%%%%%%%%%%%%%%%%%%%%%%%%%%%%%%%%%%%%%%%%%%%%

We first analyze the data obtained through the simple sampling procedure outlined above, involving $10^6$ independent disorder realizations. 

Figure \ref{fig:GOUETWsimple} shows the probability distributions of the rescaled conductance logarithm $\chi$, Eq.~\eqref{eq:chi}, for the two lead geometries considered: narrow-narrow (NN, open symbols) and wide-narrow (WN, solid symbols). Different symbols correspond to various system sizes and disorder strengths. The data are closely aligned with both the GOE and GUE Tracy-Widom distributions, represented by the blue dotted and red long-dashed lines, respectively. However, distinguishing between the distributions for the different cases is clearly challenging, as the two sets of numerical data overlap closely at this level of precision.

%%%%%%%%%%%%%%%%%%%%%%%%%%%%%%%%%%%%%%%%%%%%%%%%%%%%%%%%%%%%%%%%%%%%%%%%%%%
\begin{figure}[t]
\begin{center}
\includegraphics[height=6.5cm,width=8.5cm,angle=0]{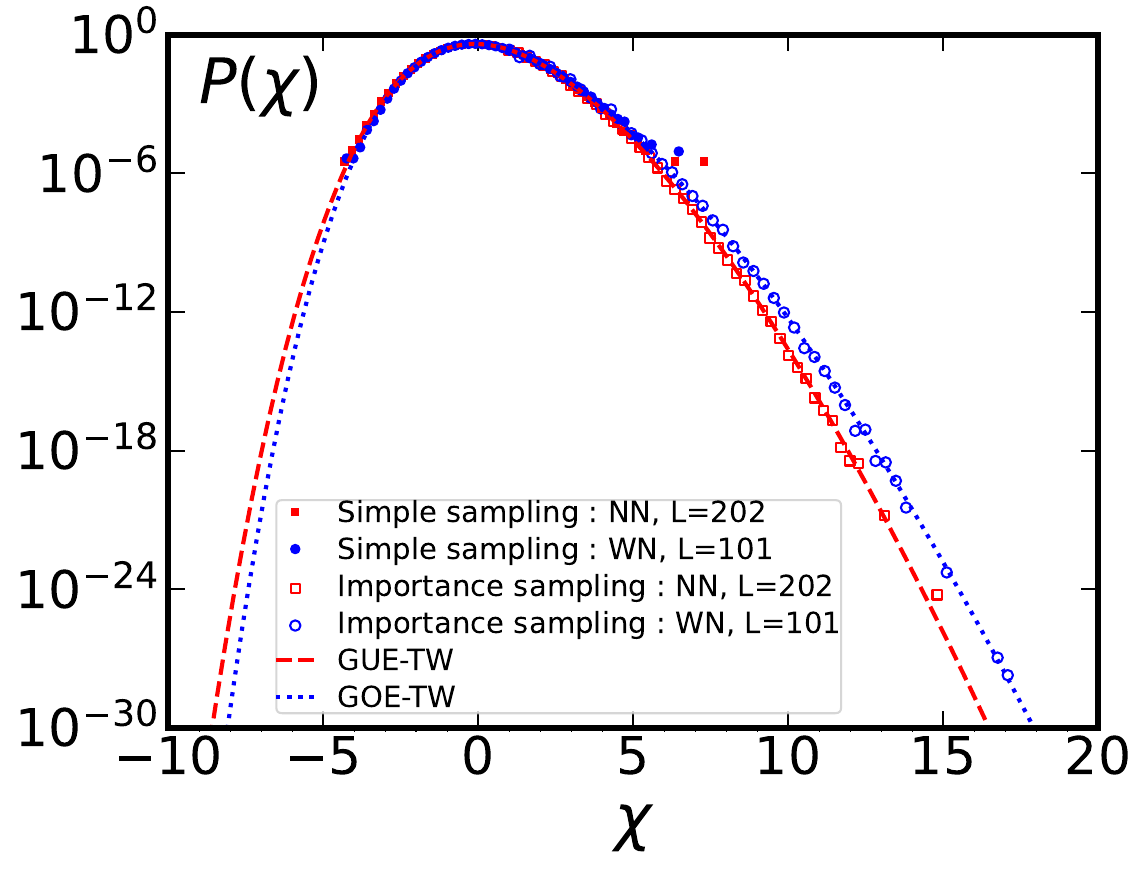}
\caption{Importance sampling results:
Distribution of the rescaled conductance logarithm \( \chi = (\ln g - \langle \ln g\rangle)/\sigma \), obtained using simple sampling (solid symbols) and importance sampling (open symbols).  
We consider the narrow-narrow (NN) and wide-narrow (WN) lead geometries, corresponding to the GUE-TW and GOE-TW classes of fluctuations, respectively, with the associated scattering regions shown in Fig.~\ref{app_fig_geometry}.  
For a uniform disorder of strength \( W=50 \), we use a scatterer of size \( L=202 \) for the NN geometry and \( L=101 \) for the WN geometry.  
Our importance sampling approach enables access to rare disorder realizations, allowing us to sample fluctuations occurring with probabilities as low as \( P(\chi) \sim 10^{-24} \) for GUE TW (red) and \( P(\chi) \sim 10^{-28} \) for GOE TW (blue).  
In this regime, the two numerical distributions become clearly distinguishable and show excellent agreement with the respective TW distributions.  
}
\label{fig:TWimpsamp}
\end{center}
\end{figure}
%%%%%%%%%%%%%%%%%%%%%%%%%%%%%%%%%%%%%%%%%%%%%%%%%%%%%%%%%%%%%%%%%%%%%%%%%%%

\subsection{Skewness in the NN and WN cases}

Another quantity that allows us to better distinguish these probability distributions 
is the skewness, which quantifies their asymmetry about the mean; see, e.g., \cite{KPZ_expt_Takeuchi2_2010}.  
In Fig.~\ref{fig_skew}, we present the behavior of the skewness of \(P(\chi)\) as a function of the disorder strength 
for different system sizes in the NN and WN cases. The skewness values were calculated using 
the simple sampling data discussed above. The results show that the skewness approaches 
the expected values for the Tracy-Widom (TW) distributions $-$ GUE for NN and GOE for WN $-$ 
as the disorder strength and system size increase. This provides an initial indication supporting the correspondence between the NN and WN lead geometries considered here and their respective TW distributions. However, this does not constitute a definitive proof, as skewness characterizes only the third moment of the distribution rather than its full shape. Consequently, the observed agreement could be coincidental, as recently demonstrated in the context of the analogy between KPZ and the 1D Heisenberg spin chain \cite{rosenberg2024dynamics}. To rigorously establish this correspondence, one must determine the full distribution and, given the similarity between GOE-TW and GUE-TW, employ methods that go beyond standard sampling.

\subsection{Importance sampling}

To definitively distinguish between the two lead configurations and classify them into the KPZ sub-universality classes, GOE TW and GUE TW, it is essential to accurately measure the tails of the probability distributions. Therefore, employing the importance sampling approach, as described in Section \ref{sec:impsamp}, is crucial.

We perform the following procedure.
The Monte Carlo (MC) process in the space of disorder configurations is guided by a function $Q(\chi)$, determined as follows: the parameters \(a\), \(b\), and \(c\) characterizing \(Q(\chi)\) (see Eq.~\eqref{eq:guiding}) are obtained by fitting \(1/Q(\chi)\) to the simple sampling distribution obtained in the previous section, in the range \(\chi \in [1, 4]\).  
For the narrow-narrow (NN) lead geometry, the fitted parameters are \(a = -0.526\), \(b = 0.915\), and \(c = 0.496\), while for the wide-narrow (WN) lead geometry, they are \(a = -0.650\), \(b = 0.868\), and \(c = 0.361\).  
Note, however, that the final results for the distribution of \(\chi\) should be independent of the guiding function, as systematically checked (see Appendix~\ref{appendix_D} and Fig.~\ref{fig_appD}).

We typically perform 10,000 MC sweeps for equilibration (where one sweep corresponds to a completely new disorder configuration of the 2D sample). In the NN case, we take measurements for the next 140,000 MC sweeps at intervals of 50 MC sweeps (since the autocorrelation time \(\tau < 50\) MC sweeps). For the WN geometry, we take measurements for the next 45,000 MC sweeps at intervals of 100 MC sweeps.  
Our importance sampling scheme enables us to explore conductance fluctuations in the range \(\chi \in [1, 15]\) for the NN geometry and \(\chi \in [1, 18]\) for the WN case.

From our MC data, we obtain the histogram \( R(\chi) \) of the stationary distribution of the above Markov process, from which we deduce the distribution \( P(\chi) = \mathcal{N} R(\chi) / Q(\chi) \), where \( \mathcal{N} \) is a normalization factor determined by fitting the simple sampling distribution. We present the obtained distributions \( P(\chi) \) in Fig.~\ref{fig:TWimpsamp}. The solid symbols represent the distributions obtained from simple sampling, whereas the open symbols correspond to the distributions from importance sampling.

It is clear that the distribution for the narrow-narrow (NN) lead configuration follows the GUE Tracy-Widom function (dashed red line), and is now clearly distinguishable from the distribution for the wide-narrow (WN) lead geometry, which follows the GOE Tracy-Widom function (dashed blue line). 
This ability to differentiate between the GUE and GOE Tracy-Widom distributions is a precision measurement that demonstrates the potential of importance sampling to resolve complex statistical behaviors that are inaccessible with simple sampling methods.

We have verified that these results are insensitive to the specific values of \( W \) and \( L \), within the strongly localized regime \( \xi \ll L \), see Appendix~\ref{appendix_E} and Fig.~\ref{fig_appE}. 
%Additionally, these findings remain consistent regardless of the disorder distribution, as detailed in the appendix. 
Therefore, we can definitively assert the presence of two distinct sub-universality classes for conductance fluctuations in two dimensions, depending on the boundary conditions, i.e., the lead geometries.

\section{Effect of a Magnetic Field on the Distribution} \label{sec:magfield}

%%%%%%%%%%%%%%%%%%%%%%%%%%%%%%%%%%%%%%%%%%%%%%%%%%%%%%%%%%%%%%%%%%%%%%%%%%%
\begin{figure}[t]
\begin{center}
\includegraphics[height=6.5cm,width=8.5cm,angle=0]{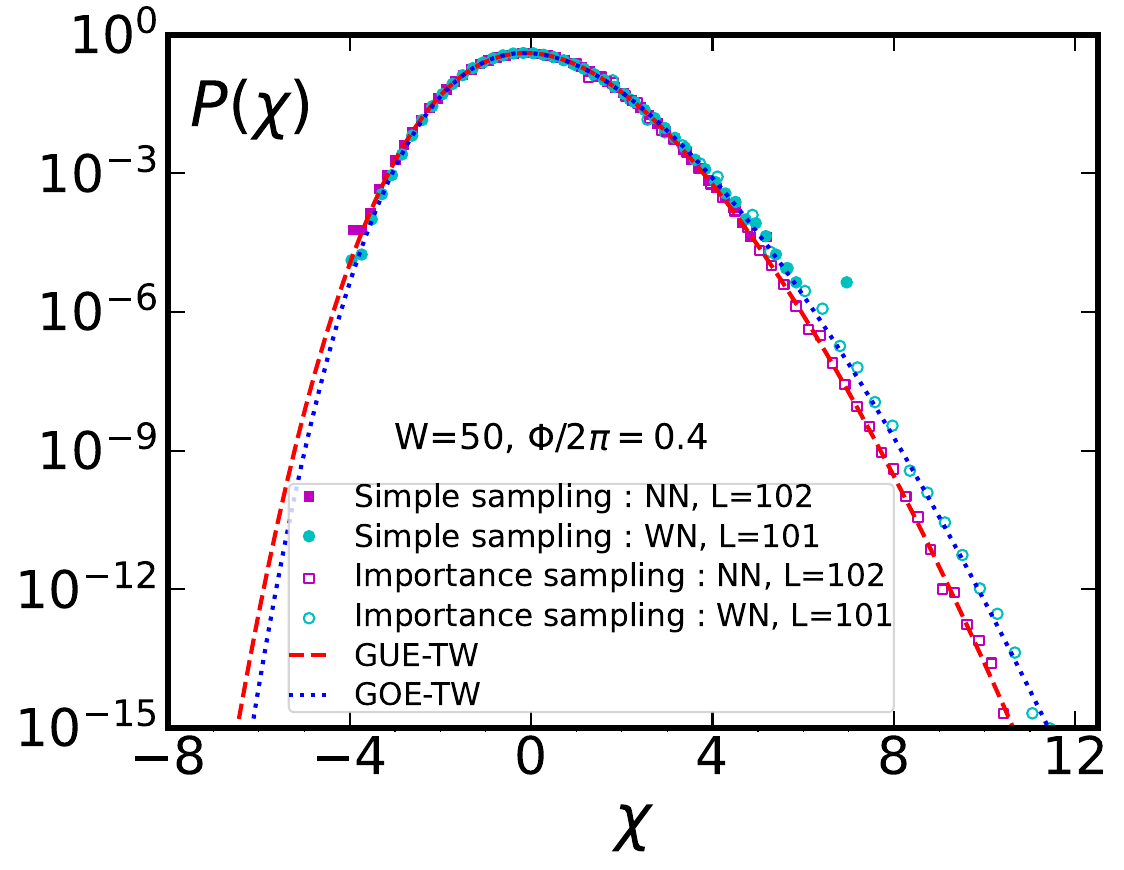}
\caption{Study of the effect of a magnetic field on importance sampling with narrow-narrow (NN) and wide-narrow (WN) leads.  
Distribution of the rescaled conductance logarithm \( P(\chi) \) obtained using importance sampling for disorder strength \( W=50 \) and a magnetic flux per unit cell \( \Phi/(2 \pi)=0.4 \). The system size is \( L=102 \) for the NN case and \( L=101 \) for the WN case. This approach allows us to probe the tail of the distribution, reaching probabilities as low as \( P(\chi) \approx 10^{-15} \) in the presence of a magnetic field.  
We observe no significant impact of the magnetic field on \( P(\chi) \), confirming its negligible influence in this regime. The data for the NN and WN cases closely follow the exact GUE-TW and GOE-TW distributions (fully consistent with the zero-field data shown in Fig.~\ref{fig:TWimpsamp}), respectively, reinforcing that universality sub-classes are determined by the lead configurations (NN or WN) rather than by time-reversal symmetry, which is controlled by the presence or absence of a magnetic field.
}
\label{fig:mag_field}
\end{center}
\end{figure}
%%%%%%%%%%%%%%%%%%%%%%%%%%%%%%%%%%%%%%%%%%%%%%%%%%%%%%%%%%%%%%%%%%%%%%%%%%%

In this section, we investigate whether these universal distributions are affected by a change in symmetry class. Specifically, we study fluctuations in the unitary symmetry class by revisiting our previous results in the presence of a uniform magnetic field characterized by \( \Phi \neq 0 \), see Eq.~\eqref{eq1}.

The application of a magnetic field disrupts time-reversal symmetry, breaking the constructive interference of time-reversed paths, thereby enhancing the localization length~\cite{Pichard_PhysRevLett.65.1812, ohtsuki1993conductance, PhysRevE.48.1764}, see appendix \ref{sec.appmag}. However, the application of a magnetic field does not affect the distribution of the rescaled conductance logarithm $\chi$, Eq.~\eqref{eq:chi}, in the strongly localized regime $L\gg \xi$. In Fig. \ref{fig:mag_field} are represented the distributions $P(\chi)$ for the NN and WN configurations, obtained from importance sampling with $\Phi/2 \pi=0.4$ and $W=50$. As in Fig.~\ref{fig:TWimpsamp} (corresponding to the absence of magnetic field), we observe that NN corresponds to GUE-TW while WN fits GOE-TW. Additional results for other disorder strengths and system sizes shown in the appendix \ref{sec.appmag} confirm this picture.

\section{Discussion}

The central question—why conductance fluctuations in the strongly localized regime of two dimensions are sensitive to boundary conditions but not to symmetry classes—can be understood through the interplay of two distinct physical mechanisms: quantum interference at length scales smaller than the localization length \(\xi\), and KPZ physics at larger scales.

This dichotomy aligns with our exact understanding of quasi-1D localization, where localized states can be decomposed as \( \psi_{\text{1D}}(r) = \phi(r) \Psi(r) \) \cite{Mirlin_PR_2000, PhysRevB.85.035109}. Here, \( \phi(r) \) exhibits rapid fluctuations at the mean free path scale, following universal random matrix theory statistics dependent on the symmetry class, whereas \( \Psi(r) \) represents the localized envelope. Under certain approximations, \( \ln \Psi(r) \) follows a diffusion equation with a drift \cite{PhysRevB.85.035109}. In this analogy, KPZ physics emerges as the diffusive fluctuations of the envelope logarithm in quasi-1D systems.

Quantum interference contributes to conductance fluctuations via universal conductance fluctuations \cite{PhysRevLett.55.1622, PhysRevB.35.1039, altshuler1985fluctuations}, leading to a conductance standard deviation of order 1. In contrast, KPZ physics governs the fluctuations of \(\ln g\), with a standard deviation scaling as \( (L/\xi)^{1/3} \). At large system sizes \( L \gg \xi \), this latter contribution dominates, making conductance fluctuations primarily influenced by KPZ physics rather than quantum interference.

However, this dominance does not extend to all observables. For instance, the sensitivity of conductance to infinitesimal disorder changes—often referred to as ``fragility" or ``chaos" in DP physics—exhibits a different behavior. As discussed in \cite{Lemarie_PRL_2019}, conductance correlations between nearly identical disorder realizations are disrupted more effectively by quantum interference than by KPZ physics.

Symmetry classes, studied here via the effect of a magnetic field breaking time-reversal symmetry, are well known to affect quantum interference contributions \cite{altshuler1994universalities, RMT_transport_Beenakker_1997, Mirlin_PR_2000}. In the DP description of Anderson localization (see Sec.~\ref{sec:FSA}), the presence of a magnetic field modifies localization properties by enhancing the localization length, describing magnetoresistance effects, but does not alter fluctuation properties \cite{Kardar_PhysRevLett.64.1816, gangopadhyay2013magnetoresistance, muller2013magnetoresistance}. In contrast, KPZ fluctuations are highly sensitive to boundary conditions \cite{KPZ_expt_Takeuchi_2010, KPZ_expt_Takeuchi2_2010, PhysRevLett.84.4882, Spohn2020kpz, Calabrese_2010, Dotsenko_2010, Prahofer_2000_PRL}, leading to the emergence of different sub-universality classes.

The deviations from the Tracy-Widom (TW) distribution observed at smaller values of $L$ or $W$ (i.e., when $\xi/L$ is not sufficiently large), as shown in Fig.~\ref{fig_skew}, can be attributed either to the competition between quantum interference and KPZ physics, or to the slow convergence of KPZ dynamics toward TW distributions. These deviations are not due to numerical inaccuracies, as our computations are exact. Additional results shown in the appendix \ref{sec.appFSS} indicate that conductance fluctuations follow KPZ scaling, \( \sigma \sim (L/\xi)^{1/3} \), in regimes where the distribution $P(\chi)$ shows deviations from the TW distribution in the tails. This suggests that KPZ physics governs the typical fluctuations (controlling $\sigma$) but the system size is insufficiently large to reach the TW distribution in the tails. Similar deviations at too short times (analogous to small system sizes) are well-documented in KPZ physics \cite{Hartmann_2018_EPL, Hartmann_2020_PRE}. A deeper understanding of this intermediate \( L \gtrsim \xi \) regime could be achieved using more advanced KPZ results, such as extreme event statistics \cite{Hartmann_2018_EPL, Hartmann_2020_PRE, LeDoussal_2016}.

\section{Summary and outlook}  
\label{sec_summ_outlook}

This study provides strong numerical evidence that conductance fluctuations in the strongly Anderson localized regime in two dimensions fall into different sub-universality classes. In all cases considered, the conductance distribution follows either the Tracy-Widom (TW) distribution for the largest eigenvalue of random matrices in the Gaussian Orthogonal Ensemble (GOE) or the Gaussian Unitary Ensemble (GUE) \cite{Tracy_Widom_1994, Tracy_Widom_1996}.

Surprisingly, applying a magnetic field breaking time-reversal symmetry does not transition the conductance distribution from GOE TW to GUE TW, challenging the paradigm that symmetry classes govern the universal properties of Anderson localization \cite{Evers_RMP_2008}. Additionally, in contrast to the Thouless criterion for localization \cite{edwards1972numerical}, another paradigm of Anderson localization, one switches between sub-universality classes by altering boundary conditions, such as modifying lead configurations in the 2D scattering system.

These paradigm shifts stem from a deep analogy between Anderson localization in 2D and KPZ physics \cite{PhysRevB.46.9984, prior2005conductance, Somoza_2007, prior2009conductance, Somoza_2015, Lemarie_PRL_2019, musen_2023}. KPZ physics \cite{KPZ_1986,Corwin_2012,KPZ_30yrs_2015,Takeuchi_2018,Spohn_2020, HALPINHEALY1995,Johansson2000,Prahofer_2000_PRL, PhysRevLett.129.260603} describes surface growth and directed polymer phenomena and was recently observed in quantum systems such as polariton condensates \cite{KPZ_expt_polariton_2022}. One of its most precise predictions is that fluctuations follow universal distributions dictated by initial conditions: the GOE TW distribution for flat initial conditions and the GUE TW distribution for droplet initial conditions \cite{KPZ_expt_Takeuchi_2010, KPZ_expt_Takeuchi2_2010, PhysRevLett.84.4882, Spohn2020kpz, Calabrese_2010, Dotsenko_2010, Prahofer_2000_PRL}.

The precision required to demonstrate this result surpasses conventional disorder sampling methods. Standard approaches cannot distinguish GOE-TW and GUE-TW distributions, as they differ significantly only in their tails. Resolving this distinction necessitates probing conductance fluctuations with a precision of \( \sim 10^{-30} \). We achieved this using an importance sampling Monte Carlo scheme for disorder configurations \cite{MC_book_Newman,MC_book_Landau, Monthus_2006,Saito_2010,Hartmann_2018_EPL,Biroli_2022}, enabling us to access events as rare as one in \( 10^{30} \) realizations and providing conclusive evidence for KPZ physics in the localized regime.

The theoretical understanding of KPZ physics in (1+1) dimensions has reached an exceptional level of mathematical and physical rigor, yielding numerous universal analytical predictions. These insights may provide a fresh perspective on the relatively less understood physics of Anderson localization in two dimensions. For instance, beyond the GOE and GUE subclasses, other fundamental KPZ sub-universality class—the Baik-Rains distribution, associated with the stationary initial condition \cite{PhysRevLett.108.190603, borodin2015height, PhysRevLett.124.250602} and the GSE TW distribution \cite{Somoza_2015}—remain to be explored in the context of 2D Anderson localization. Our study may also offer guidance for advancing traditional approaches to Anderson localization, whether through field-theoretic methods~\cite{Evers_RMP_2008} or generalized Dorokhov-Mello-Pereyra-Kumar (DMPK) formulations~\cite{dorokhov1982transmission, mello1988macroscopic, RevModPhys.69.731, muttalib2002generalization, PhysRevB.72.125317, douglas2014generalized, PhysRevB.106.184203}.

More broadly, the DP physics underlying Anderson localization in 2D—and the associated rich KPZ phenomenology—should extend, based on perturbative arguments in the strong disorder regime (see, e.g., \cite{FIM:PRB10, Monthus_2012, lemarie2013universal, PhysRevB.93.054201}), to a wide class of disorder-induced localized phases. Notable examples include the MBL phase \cite{PhysRevB.102.014208, biroli2017delocalized} and the Anderson transition in random graphs \cite{biroli2012difference, PhysRevLett.118.166801, PhysRevResearch.2.012020, PhysRevB.106.214202, kravtsov2017non, Biroli_2022, PhysRevB.110.184210}, where this analogy has already been leveraged to characterize these transitions and their non-ergodic properties. In such settings, the importance sampling method used in our study—which effectively identifies rare disorder configurations that play a significant role—should prove particularly valuable (see \cite{Biroli_2022, PhysRevB.110.014205} for pioneering applications). This is especially relevant in the context of many-body localization, where the impact of rare thermal inclusions, induced by specific disorder configurations, on the stability of an otherwise localized phase has emerged as a crucial yet numerically challenging question \cite{sierant2024many}.

\begin{acknowledgements} 

We thank M. Sen, J. Gong, N. Izem, B. Georgeot, P. Le Doussal, G. Schehr, A. M. Somoza, M. Ortu\~no, and J. Prior, for helpful discussions.
We acknowledge the use of computational resources at the Singapore National Super Computing Centre (NSCC) ASPIRE-2A cluster and at Calcul en Midi-Pyrénées (CALMIP) for our simulations. 
This work is supported by the Singapore Ministry of Education Academic Research Funds Tier II (MOE-T2EP50223-0009 and MOE-T2EP50222-0005), the Singapore National Research Foundation Investigator Award (NRF-NRFI06-2020-0003), and the ANR projects Gladys (ANR-19-CE30-0013), QUTISYM (ANR-23-PETQ-0002) and ManyBodyNet, the EUR Grant NanoX No. ANR-17-EURE-0009.

\end{acknowledgements}

\newpage

\appendix

%\label{append}

\section{Lead configurations $NN$ and $WN$}

%%%%%%%%%%%%%%%%%%%%%%%%%%%%%%%%%%%%%%%%%%%%%%%%%%%%%%%%%%%%%%%%%%%%%%%%%%%
\begin{figure}
\setcounter{figure}{0}
\renewcommand{\thefigure}{A\arabic{figure}}
\centerline{
\includegraphics[height=4.5cm,width=8.0cm,angle=0]{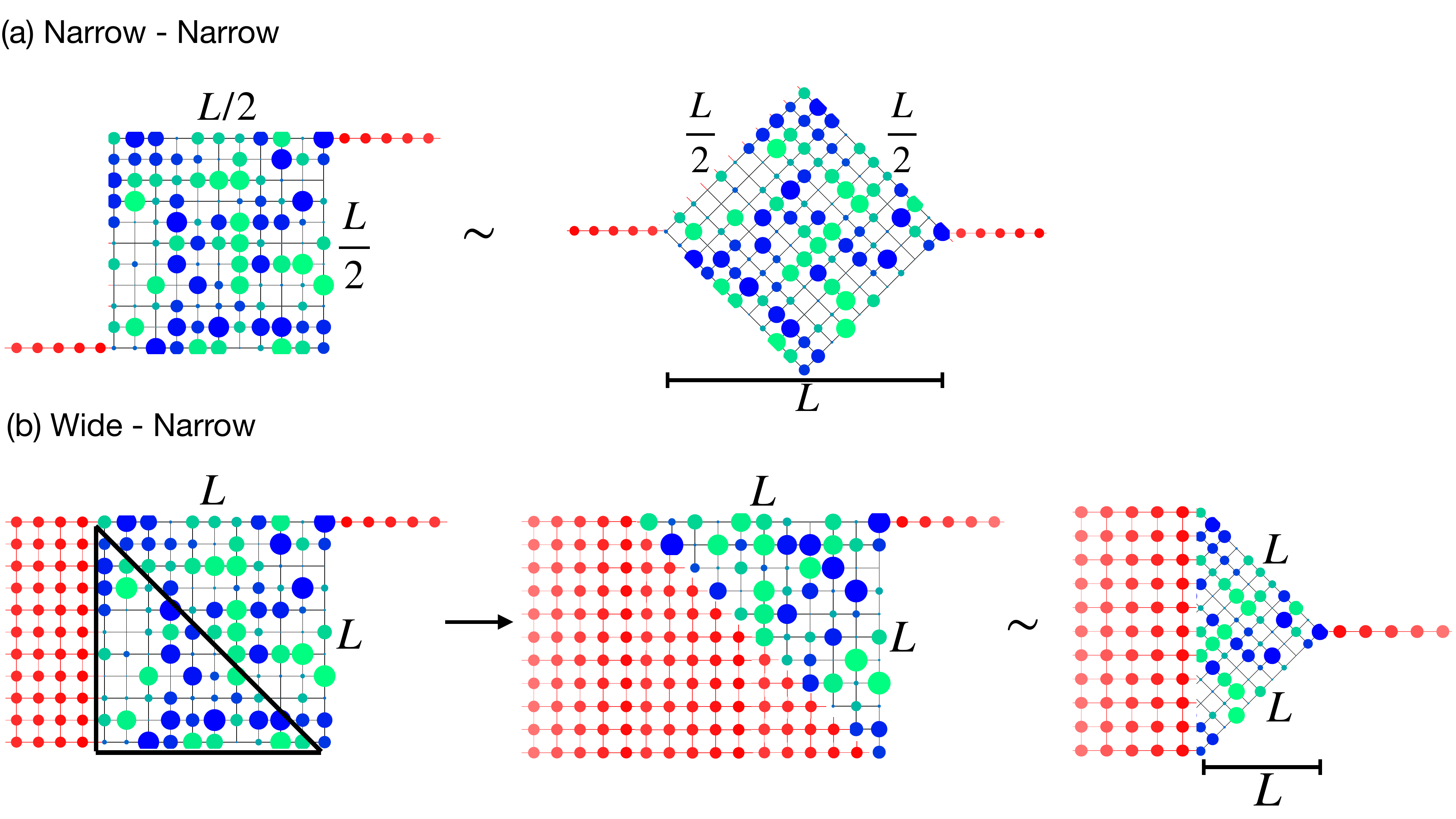}
}
\caption{
(i) The `narrow-narrow' (NN) configuration consists of a disordered scattering system of size \(L/2 \times L/2\), with 1D perfect leads attached at the bottom-left and top-right corners.  
(ii) The `wide-narrow' (WN) configuration consists of a scattering system of size \(L \times L\). A wide lead, matching the width of the scatterer, is attached to the left edge, while a 1D lead is connected to the top-right corner. The disorder is absent in the bottom triangular region of the scatterer, making it identical to the left lead.  
}
\label{app_fig_geometry}
\end{figure}
%%%%%%%%%%%%%%%%%%%%%%%%%%%%%%%%%%%%%%%%%%%%%%%%%%%%%%%%%%%%%%%%%%%%%%%%%%%

{\it Narrow-Narrow (NN) Configuration:}  
For this configuration, the scattering system has a size of \( L/2 \times L/2 \), with the left 1D lead attached at the bottom-left corner and the right 1D lead attached at the top-right corner, as illustrated in Figure \ref{app_fig_geometry} (top row).  

{\it Wide-Narrow (WN) Configuration:}  
This configuration features an \(L \times L\) scattering system. A left lead, matching the width of the scattering region, is attached to the left edge, while a 1D lead is connected to the top-right corner. The disorder is absent in the bottom triangular region of the scatterer, making it identical to the left lead. This is illustrated in Figure \ref{app_fig_geometry} (bottom row).

%%%%%%%%%%%%%%%%%%%%%%%%%%%%%%%%%%%%%%%%%%%%%%%%%%%%%%%%%%%%%%%%%%%%%%%%%%%
\begin{figure*}
\setcounter{figure}{1}
\renewcommand{\thefigure}{A\arabic{figure}}
\begin{center}
\includegraphics[width=0.95\linewidth,angle=0]{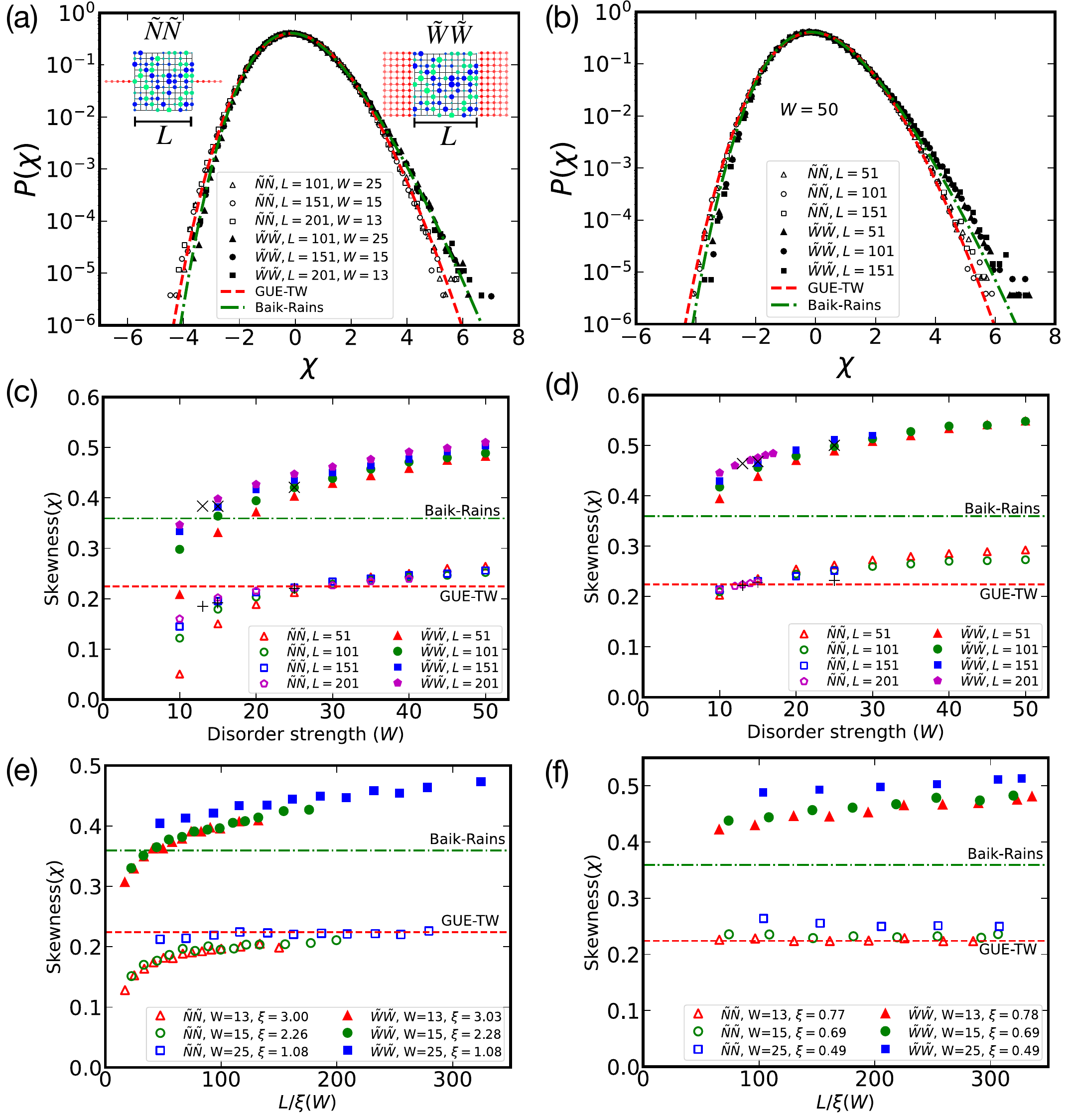}
\caption{%
Nuremical results for the scatterer and lead geometries studied by Somoza \textit{et al.}~\cite{prior2005conductance, Somoza_2007, prior2009conductance}.(a)~Distribution of the rescaled conductance logarithm, 
\(\chi = (\ln g - \langle \ln g \rangle)/\sigma\).
Inset: the setups considered in \cite{prior2005conductance, Somoza_2007, prior2009conductance} with a square scatterer and either narrow-narrow ($\tilde{N}\tilde{N}$) or wide-wide ($\tilde{W}\tilde{W}$) leads, distinct from our configurations in Fig.~\ref{fig1}. 
Using the parameters considered in \cite{Somoza_2007} ($L \times L$ scatterers with \(L = 101, 151, 201\); uniform disorder strengths \(W = 25, 15, 13\); $10^6$ disorder realizations per case), we find that the $\tilde{N}\tilde{N}$ geometry follows the GUE-TW distribution (red dashed line), while $\tilde{W}\tilde{W}$ aligns with the Baik-Rains distribution (green dot-dashed line). 
(b)~However, for stronger disorder (\(W = 50\)) and \(L = 51, 101, 151\), distributions deviate significantly from both GUE-TW and Baik-Rains.
(c,d)~Skewness as a function of disorder strength for $\tilde{N}\tilde{N}$ (empty symbols) and $\tilde{W}\tilde{W}$ (solid symbols), with uniform (c) and Gaussian (d) disorder. 
Skewness is computed via simple sampling over $10^6$ disorder realizations. 
Dashed red and dot-dashed green lines mark GUE-TW and Baik-Rains skewness values, respectively. 
Crosses and plus signs indicate data from~\cite{Somoza_2007}. 
In both cases, deviations grow with disorder and depend on the disorder distribution.
(e,f)~Skewness versus \(L/\xi\) for $\tilde{N}\tilde{N}$ (empty) and $\tilde{W}\tilde{W}$ (solid) leads at moderate disorder strengths \(W = 13, 15, 25\) considered in \cite{Somoza_2007}, for uniform (e) and Gaussian (f) disorder. 
For $\tilde{N}\tilde{N}$, skewness converges to GUE-TW with increasing \(L/\xi\), consistent with the fact that network structure should play a minor role when $\xi>1$. 
In contrast, $\tilde{W}\tilde{W}$ shows systematic deviation from Baik-Rains even at moderate disorder, contradicting the universality claim of~\cite{prior2005conductance, Somoza_2007, prior2009conductance}.
}
\label{fig:SOPTW}
\end{center}
\end{figure*}
%%%%%%%%%%%%%%%%%%%%%%%%%%%%%%%%%%%%%%%%%%%%%%%%%%%%%%%%%%%%%%%%%%%%%%%%%%%

\section{Numerical results on the configurations \(\tilde{N}\tilde{N}\) and \(\tilde{W}\tilde{W}\) considered in previous studies.} \label{sec.appSOP}

In this section, we present the numerical results obtained for the \(\tilde{N}\tilde{N}\) and \(\tilde{W}\tilde{W}\) configurations, which were previously investigated in studies on conductance fluctuations in the 2D Anderson localized regime \cite{prior2005conductance, Somoza_2007, prior2009conductance}. Our analysis extends these earlier works by considering a wider range of disorder strengths \(W\) and system sizes \(L\).  

These lead configurations differ fundamentally from those considered in the present study:  
(i) In \(\tilde{N}\tilde{N}\), the 1D leads are attached at the midpoints of opposite sides of the square sample, whereas in our setup (Fig.~\ref{fig1}), they are positioned at opposite corners.  
(ii) In \(\tilde{W}\tilde{W}\), the leads extend across the entire width of the square sample and are connected to opposite sides, with periodic boundary conditions imposed in the direction perpendicular to the leads.

The results are summarized in Fig.~\ref{fig:SOPTW}. 
While we confirm the findings of~\cite{prior2005conductance, Somoza_2007, prior2009conductance} using their original parameters $W$ and $L$ (panel~(a)), we observe significant deviations at strong disorder (panel~(b)). 
A systematic study of the skewness across a broad range of disorder strengths ($W = 10$ to $50$) and system sizes ($L = 50$ to $400$) reveals clear deviations at strong disorder for the $\tilde{N}\tilde{N}$ geometry, and across all disorder strengths for the $\tilde{W}\tilde{W}$ geometry (panels~(c) and~(d)), with the deviations dependent on the disorder distribution. 
In other words, while the results of~\cite{prior2005conductance, Somoza_2007, prior2009conductance} are confirmed for $\tilde{N}\tilde{N}$ leads at moderate disorder and large system sizes ($L \gg \xi > 1$), we find significant deviations for $\tilde{W}\tilde{W}$ leads at all disorder strengths and large $L$ (panels~(e) and~(f)).

These discrepancies indicate that the \(\tilde{W}\tilde{W}\) configuration does not fall within the stationary/Baik-Rains universality class of KPZ. This supports our reservations regarding the boundary conditions used, which differ from the stationary KPZ distribution, as well as the observable considered, which is not equivalent to the height difference \(h(x,t) - h(x,0)\). The \(\tilde{N}\tilde{N}\) case is less problematic, as deviations appear only in the strong disorder regime \(\xi < 1\), where the network structure plays a significant role. In this regime, deviations for \(\tilde{N}\tilde{N}\) arise because not all directed paths from the left to right leads have the same length—an essential requirement for observing the GUE-TW distribution. At moderate disorder \(\xi > 1\), the influence of the network geometry should be negligible, consistent with our observation of GUE-TW scaling at large system sizes, in agreement with~\cite{prior2005conductance, Somoza_2007, prior2009conductance}.

%%%%%%%%%%%%%%%%%%%%%%%%%%%%%%%%%%%%%%%%%%%%%%%%%%%%%%%%%%%%%%%%%%%%%%%%%%%
\begin{figure}[t]
\setcounter{figure}{2}
\renewcommand{\thefigure}{A\arabic{figure}}
\begin{center}
\includegraphics[width=0.98\linewidth,angle=0]{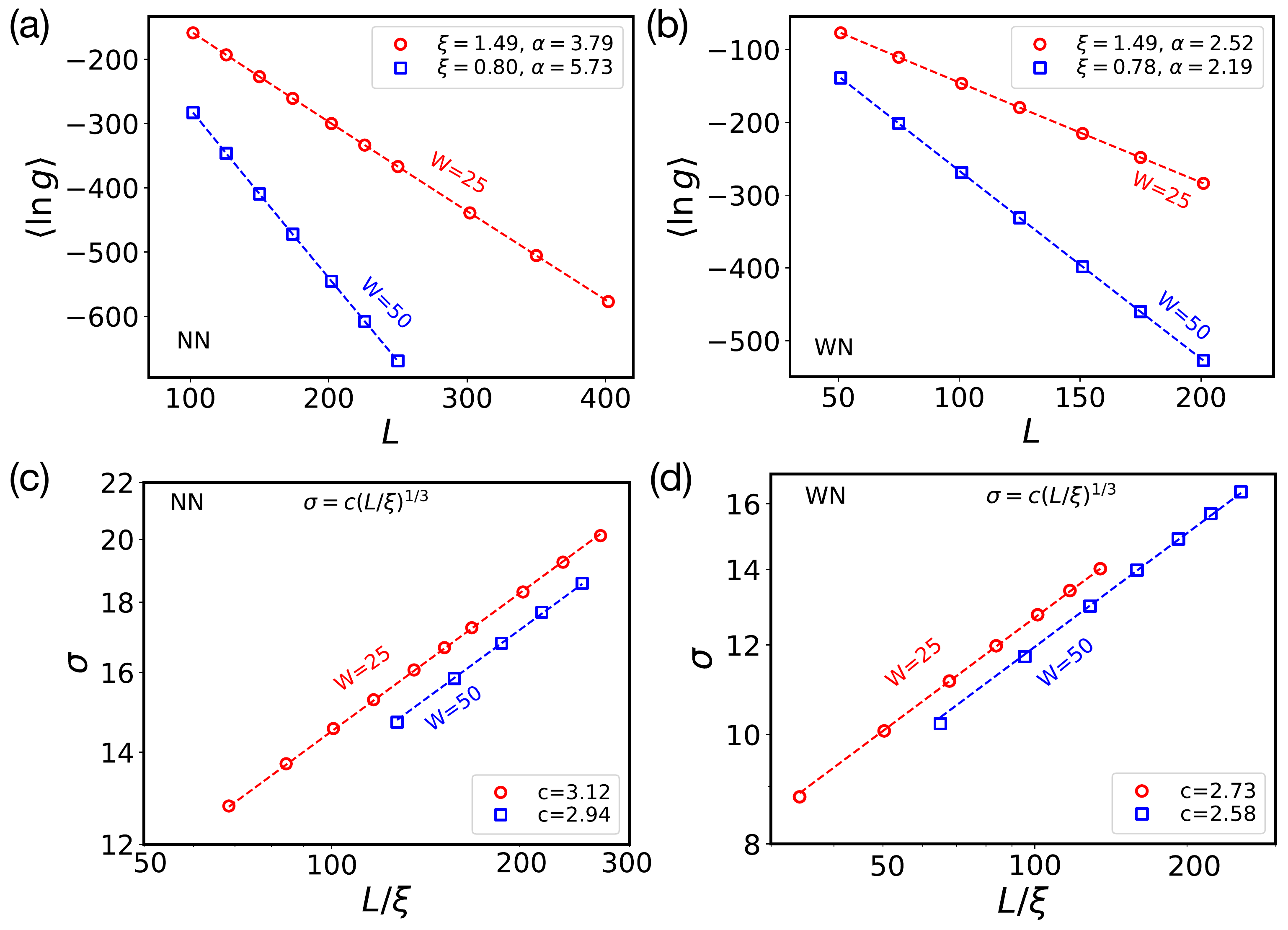}
\caption{
(a)-(b)~Variation of the average conductance logarithm, \(\langle \ln g \rangle\), for narrow-narrow (NN) and wide-narrow (WN) leads as a function of the system size \(L\) of the scattering region, for uniform disorder values \(W=25, 50\) and $\Phi=0$. The data points are obtained from our computations, while the dashed lines represent fits to  $
\langle \ln g \rangle = c -\frac{2L}{\xi} + \alpha \left(\frac{L}{\xi}\right)^{1/3} \langle \chi \rangle$.  
We use \(\langle \chi \rangle = 1.7710868074116026\) for NN and \(\langle \chi \rangle = 1.2065335745820245\) for WN, as estimated from the theoretical values of the GUE and GOE Tracy-Widom distributions~\cite{KPZ_data_Prahofer_Spohn}.  
(c)-(d)~Standard deviation of the conductance logarithm,  
$
\sigma = \sqrt{\langle(\ln g - \langle \ln g \rangle)^2 \rangle},
$  
as a function of the system size \(L\), for uniform disorder values \(W=25, 50\). The data points are obtained from our computations, while the dashed lines represent fits to  
$
\sigma = c \left(\frac{L}{\xi}\right)^{1/3}.
$  
}

\label{fig_app_scaling}
\end{center}
\end{figure}
%%%%%%%%%%%%%%%%%%%%%%%%%%%%%%%%%%%%%%%%%%%%%%%%%%%%%%%%%%%%%%%%%%%%%%%%%%%

\section{Additional results via simple sampling procedure}

\subsection{KPZ scaling properties}\label{sec.appFSS}

%%%%%%%%%%%%%%%%%%%%%%%%%%%%%%%%%%%%%%%%%%%%%%%%%%%%%%%%%%%%%%%%%%%%%%%%%%%
\begin{figure}
\setcounter{figure}{3}
\renewcommand{\thefigure}{A\arabic{figure}}
\begin{center}
\includegraphics[width=0.98\linewidth,angle=0]{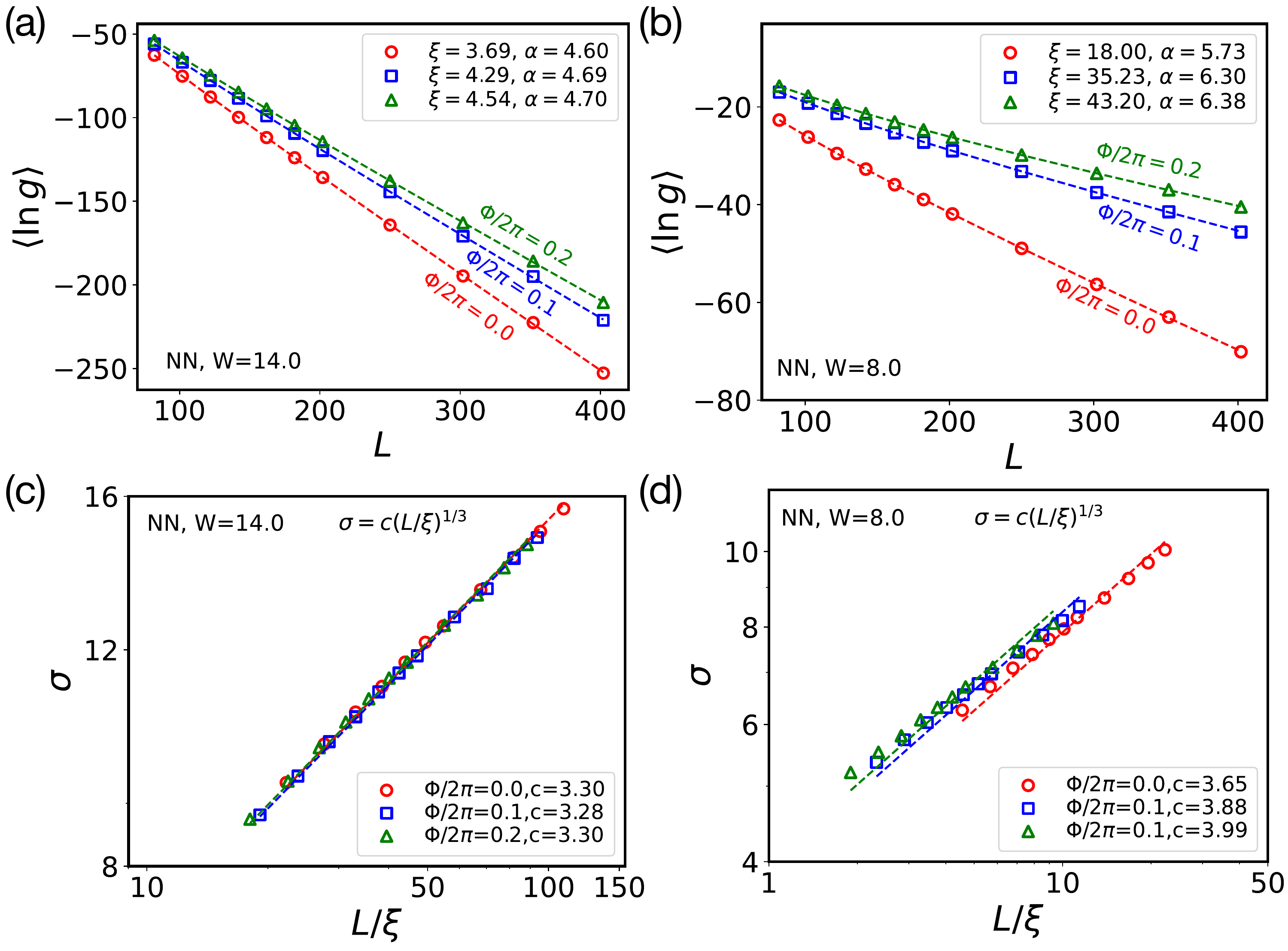}
\caption{KPZ scaling behavior in the presence of an external magnetic field with the narrow-narrow (NN) lead geometry, for uniform disorder values \(W=14, 8\).  
(a)-(b)~Variation of \(\langle \ln g \rangle\) as a function of the system size \(L\) of the scattering region, for magnetic flux values \(\Phi/2\pi=0.0, 0.1\), and \(0.2\). The data points are obtained from our computations, while the dashed line represents the fit to  
\(
\langle \ln g \rangle = c -\frac{2L}{\xi} + \alpha \left(\frac{L}{\xi}\right)^{1/3} \langle \chi \rangle.
\) 
We use \(\langle \chi \rangle = 1.7710868074116026\), as estimated from the theoretical values of the GUE Tracy-Widom distribution~\cite{KPZ_data_Prahofer_Spohn}.  
(c)-(d)~Standard deviation of the conductance logarithm,  
\(
\sigma = \sqrt{\langle(\ln g - \langle \ln g \rangle)^2 \rangle},
\) 
as a function of the system size \(L\), for magnetic flux values \(\Phi/2\pi=0.0, 0.1\), and \(0.2\). The data points are obtained from our computations, while the dashed line represents the fit to  
\(
\sigma = c \left(\frac{L}{\xi}\right)^{1/3}.
\)  
}
\label{fig_app_scaling2}
\end{center}
\end{figure}
%%%%%%%%%%%%%%%%%%%%%%%%%%%%%%%%%%%%%%%%%%%%%%%%%%%%%%%%%%%%%%%%%%%%%%%%%%%

Drawing on the analogy with directed polymers (DP) and Kardar-Parisi-Zhang (KPZ) physics, the logarithm of the conductance exhibits a non-trivial finite-size scaling behavior in the strongly localized regime (\(L \gg \xi\)); see \cite{PhysRevB.46.9984, prior2005conductance, Somoza_2007, prior2009conductance, Somoza_2015, Lemarie_PRL_2019, musen_2023} for previous studies. Specifically, we observe the following scaling relations in Figs.~\ref{fig_app_scaling} and \ref{fig_app_scaling2}:  
$(i)$~ For the disorder-averaged conductance logarithm:  
\[
\langle \ln g \rangle \approx -\frac{2L}{\xi} + \alpha \left(\frac{L}{\xi}\right)^{1/3} \langle \chi \rangle,
\]
where the first term represents the expected exponential decay of the conductance, governed by the localization length \(\xi\), while the second term captures the leading-order fluctuations. Here, \(\alpha\) is a constant prefactor that sets the fluctuation scale.  
$(ii)$~ Scaling of the standard deviation of the conductance logarithm:  
\[
\sigma= \sqrt{\langle(\ln g - \langle \ln g \rangle)^2 \rangle} \propto L^{1/3},
\]
which reflects the universal growth of fluctuations in this regime.  
Importantly, these scaling properties are \textit{independent} of the lead configurations and the time-reversal symmetry breaking induced by the presence of a magnetic field.

%%%%%%%%%%%%%%%%%%%%%%%%%%%%%%%%%%%%%%%%%%%%%%%%%%%%%%%%%%%%%%%%%%%%%%%%%%%
\begin{figure}[t]
\setcounter{figure}{4}
\renewcommand{\thefigure}{A\arabic{figure}}
\centerline{
\includegraphics[height=6.0cm,width=8.0cm,angle=0]{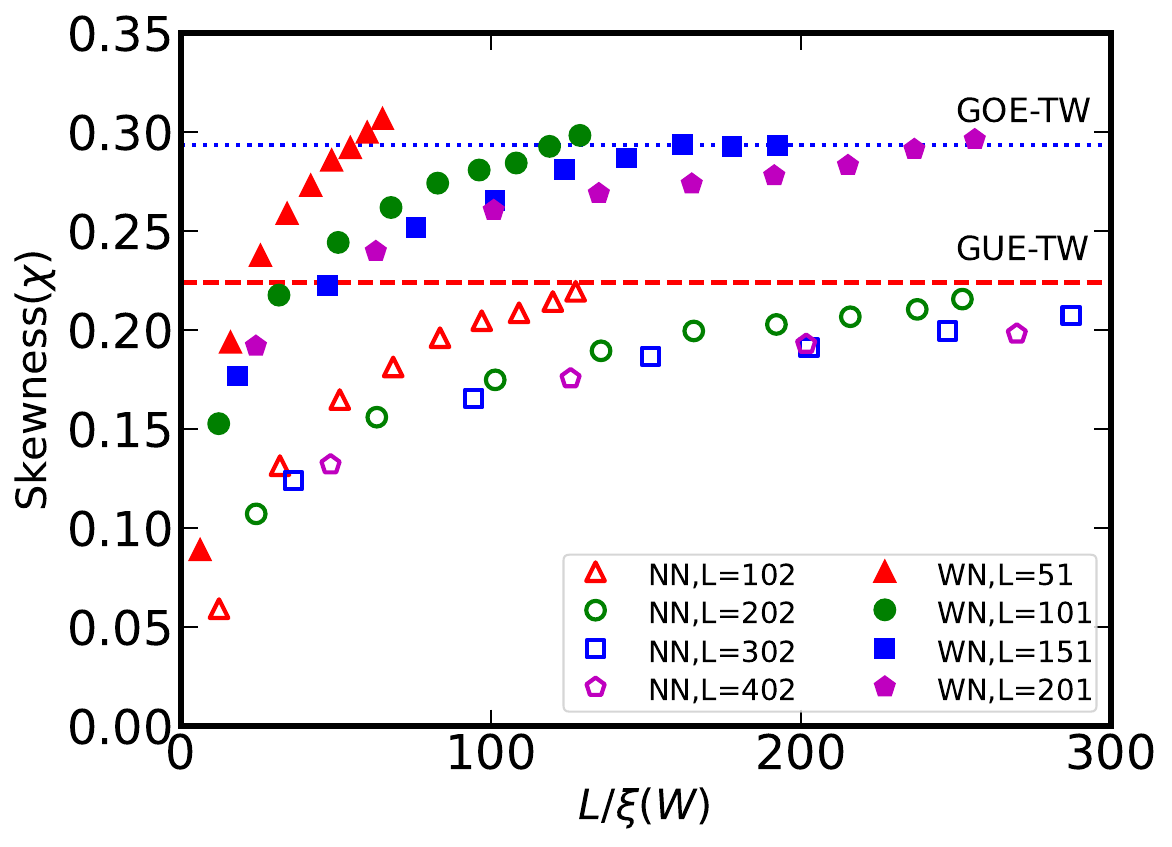}
}
\caption{
Dependence of the skewness of \( P(\chi) \), shown in Fig.~\ref{fig_skew}, on \( L/\xi \) for the NN lead geometry (empty data points) and the WN lead geometry (filled data points). The skewness is computed using simple sampling with \( 10^6 \) independent disorder realizations. The dashed red line and the dotted blue line represent the skewness values of the GUE and GOE Tracy-Widom (TW) distributions, respectively.
}
\label{fig_skew2}
\end{figure}
%%%%%%%%%%%%%%%%%%%%%%%%%%%%%%%%%%%%%%%%%%%%%%%%%%%%%%%%%%%%%%%%%%%%%%%%%%%

\subsection{Finite-size scaling of the skewness}

In Figure~\ref{fig_skew2}, we present the skewness of the distribution of the rescaled conductance logarithm, \(P(\chi)\), shown in Figure~\ref{fig_skew}, as a function of \(L/\xi\) for both narrow-narrow (NN) and wide-narrow (WN) lead geometries. The computation is based on simple sampling with \(10^6\) independent disorder realizations.  
As seen in Figure~\ref{fig_skew2}, our results indicate that for \(L/\xi \gg 1\), the skewness converges to the expected values of the Tracy-Widom (TW) distributions--GUE-TW for NN leads and GOE-TW for WN leads.

\section{Additional Results from the Importance Sampling Procedure}

\subsection{Independence of Importance Sampling Results from the Choice of Guiding Function}
\label{appendix_D}

%%%%%%%%%%%%%%%%%%%%%%%%%%%%%%%%%%%%%%%%%%%%%%%%%%%%%%%%%%%%%%%%%%%%%%%%%%%
\begin{figure}[t]
\setcounter{figure}{5}
\renewcommand{\thefigure}{A\arabic{figure}}
\centerline{
\includegraphics[height=6.0cm,width=8.0cm,angle=0]{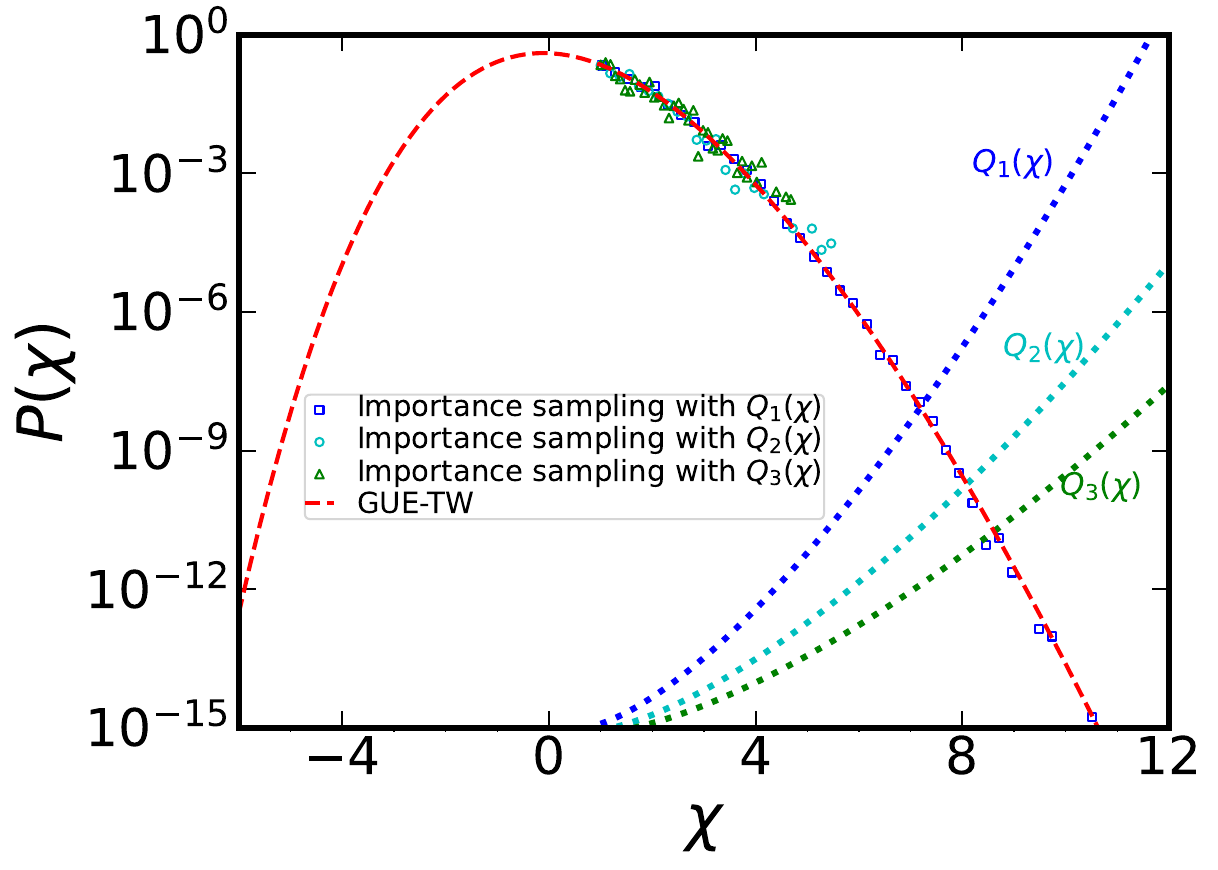}
}
\caption{Independence of \(P(\chi)\) with respect to the choice of the guiding function in the importance sampling procedure. We consider the narrow-narrow (NN) lead geometry with uniform disorder strength \(W=50\) and a system size of \(L=202\). The guiding function \( Q(\chi) = e^{-a +b|\chi|^{\eta} - c \ln|\chi|} \) is characterized by \(\eta= -3/2\) and three parameters: \(a\), \(b\), and \(c\). The three different guiding functions, \(Q_1(\chi)\), \(Q_2(\chi)\), and \(Q_3(\chi)\), are represented by dotted lines. We use the same values of \(a=-0.52649624\) and \(c=0.49634386\) for \(Q_1(\chi)\), \(Q_2(\chi)\), and \(Q_3(\chi)\), while setting \(b=0.91554331\) for \(Q_1(\chi)\), \(b=0.60425858\) for \(Q_2(\chi)\), and \(b=0.45771655\) for \(Q_3(\chi)\). Our results show that importance sampling produces disorder realizations with \(\chi\) values that follow the GUE Tracy-Widom (GUE-TW) distribution, irrespective of the chosen guiding function \(Q_1(\chi)\), \(Q_2(\chi)\), or \(Q_3(\chi)\). However, an appropriately chosen guiding function—such as \(Q_1(\chi) \approx 1/P(\chi)\)—greatly enhances the efficiency of sampling the tail of the distribution compared to alternatives like \(Q_2(\chi)\) and \(Q_3(\chi)\).  
}
\label{fig_appD}
\end{figure}
%%%%%%%%%%%%%%%%%%%%%%%%%%%%%%%%%%%%%%%%%%%%%%%%%%%%%%%%%%%%%%%%%%%%%%%%%%%

We demonstrate that the choice of a particular guiding function in our importance sampling scheme does not affect the shape of the distribution \(P(\chi)\).  
In Figure~\ref{fig_appD}, we compare three different guiding functions, \(Q_1(\chi)\), \(Q_2(\chi)\), and \(Q_3(\chi)\), each characterized by distinct sets of parameters \(a, b,\) and \(c\) in Eq.~\eqref{eq:guiding}. The distributions of the rescaled conductance logarithm, \(P(\chi)\), obtained using our importance sampling procedure for the narrow-narrow lead geometry, all coincide with the GUE Tracy-Widom distribution, irrespective of the guiding function used. However, an appropriately chosen guiding function--such as \(Q_1(\chi) \approx 1/P(\chi)\)--greatly enhances the efficiency of sampling the tail of the distribution compared to alternatives like \(Q_2(\chi)\) and \(Q_3(\chi)\).

%%%%%%%%%%%%%%%%%%%%%%%%%%%%%%%%%%%%%%%%%%%%%%%%%%%%%%%%%%%%%%%%%%%%%%%%%%%
\begin{figure}[ht!]
\setcounter{figure}{6}
\renewcommand{\thefigure}{A\arabic{figure}}
\centerline{
\includegraphics[height=6.0cm,angle=0]{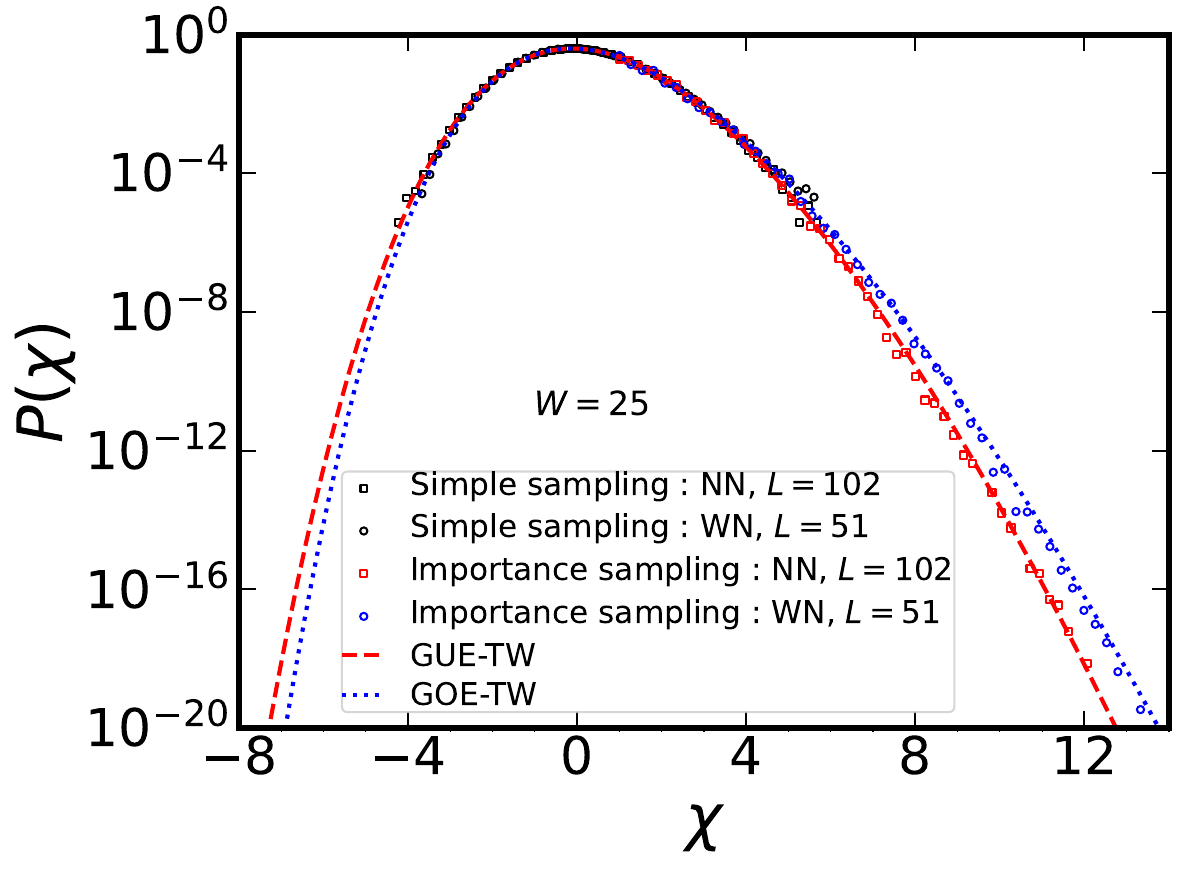}
}
\caption{
Distribution of \(\chi\), \(P(\chi)\), obtained using importance sampling for Gaussian-distributed disorder with zero mean and standard deviation \(W=25\). The system size is \(L=102\) for the NN configuration (red square symbols) and \(L=51\) for the WN configuration (blue circle symbols). The agreement with the respective Tracy-Widom distributions--GUE-TW (red dashed line) and GOE-TW (blue dashed line)--is evident down to \(10^{-20}\), highlighting the universality of this classification.  
}
\label{fig_appE}
\end{figure}
%%%%%%%%%%%%%%%%%%%%%%%%%%%%%%%%%%%%%%%%%%%%%%%%%%%%%%%%%%%%%%%%%%%%%%%%%%%

%%%%%%%%%%%%%%%%%%%%%%%%%%%%%%%%%%%%%%%%%%%%%%%%%%%%%%%%%%%%%%%%%%%%%%%%%%%
\begin{figure*}
\setcounter{figure}{7}
\renewcommand{\thefigure}{A\arabic{figure}}
\centerline{
\includegraphics[height=4.5cm,width=16.5cm,angle=0]{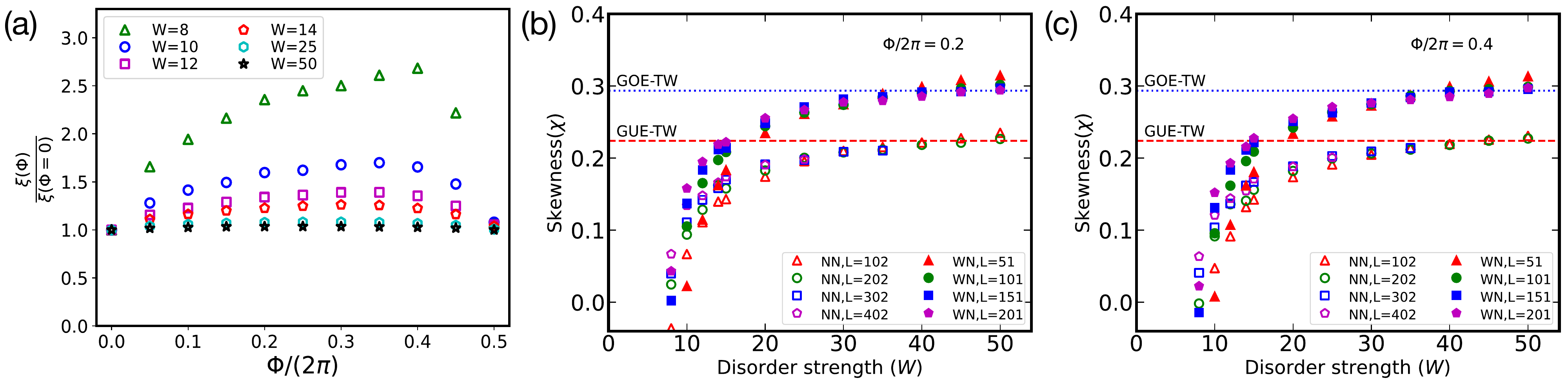}
}
\caption{Effect of a magnetic field on the localization length and the skewness of \(P(\chi)\). 
(a)~Ratio of localization lengths \( \xi(\Phi)/\xi(\Phi=0) \) as a function of magnetic flux per unit cell \( \Phi \). The magnetic field breaks time-reversal symmetry, suppressing the constructive interference of time-reversed paths and thereby enhancing the localization length \( \xi(\Phi) \). This enhancement is expected to reach a factor of two when \( \xi(\Phi=0) \) is large~\cite{Pichard_PhysRevLett.65.1812, ohtsuki1993conductance, PhysRevE.48.1764}, as observed for \( W=8 \). For stronger disorder, such as \( W=50 \), where \( \xi(\Phi=0) \) is on the order of the lattice spacing, the enhancement is significantly reduced. 
(b)-(c)~Study of finite-size effects on the skewness of the distribution of the rescaled conductance logarithm \( \chi \), as a function of disorder strength, for narrow-narrow (NN) and wide-narrow (WN) lead geometries in the presence of a magnetic field. We show the evolution of skewness with disorder and system size for magnetic flux values \( \Phi/2\pi = 0.2 \) (panel b) and \( \Phi/2\pi = 0.4 \) (panel c). The red dashed and blue dotted lines indicate the skewness values of the GUE and GOE Tracy-Widom (TW) distributions, respectively. The results confirm that, even with broken time-reversal symmetry, the skewness converges to the expected TW values in the limit of large disorder and system size, depending on the lead geometry.
}
\label{fig_skw_withfield}
\end{figure*}
%%%%%%%%%%%%%%%%%%%%%%%%%%%%%%%%%%%%%%%%%%%%%%%%%%%%%%%%%%%%%%%%%%%%%%%%%%%

\subsection{Gaussian Distributed Disorder}
\label{appendix_E}

In this section, we present additional numerical results based on Gaussian-distributed disorder. In Figure~\ref{fig_appE}, we show the distributions \(P(\chi)\) obtained using importance sampling for NN and WN configurations, which exhibit perfect agreement with the respective GUE-TW and GOE-TW distributions. This insensitivity to the disorder distribution serves as a clear signature of the universality of this classification.

\section{Additional Results on the Effects of a Magnetic Field} \label{sec.appmag}

In Fig.~\ref{fig_skw_withfield}(a), we first examine the impact of a uniform magnetic field on the localization length \( \xi(\Phi) \).  
As predicted in~\cite{Pichard_PhysRevLett.65.1812, ohtsuki1993conductance, PhysRevE.48.1764}, this enhancement is expected to reach a universal factor of two when \( \xi(\Phi=0) \) is large. Indeed, for weak disorder (\( W=8 \)), we observe a peak value of \( \xi(\Phi)/\xi(\Phi=0) \approx 2.68 \) at \( \Phi/2\pi = 0.4 \). In contrast, for stronger disorder (\( W=50 \)), where \( \xi(\Phi=0) \) is much smaller (on the order of the lattice spacing), the enhancement is significantly reduced, with a peak value of \( \xi(\Phi)/\xi(\Phi=0) \approx 1.04 \) at \( \Phi/2\pi = 0.25 \).  

Figures~\ref{fig_skw_withfield}(b) and \ref{fig_skw_withfield}(c) show the skewness of \( P(\chi) \) as a function of disorder strength for the NN and WN lead configurations, under magnetic flux \( \Phi/2\pi = 0.2 \) (panel (b)) and \( \Phi/2\pi = 0.4 \) (panel (c)).  
We observe that even in the presence of a magnetic field, the skewness converges, in the limits of large disorder and large system size, to the expected values of the GUE and GOE Tracy-Widom distributions, for the NN and WN configurations respectively.

\newpage         
\bibliography{ref}

\end{document}